\documentclass[a4paper,USenglish]{article}

\usepackage{thm-restate}
\usepackage{amsmath,amssymb,amsthm}
\usepackage{thmtools,thm-restate}
\usepackage{comment}
\usepackage{a4wide}
\usepackage[numbers,sort]{natbib}
\usepackage{tikz}
\usetikzlibrary{arrows.meta,positioning,shapes,calc,decorations.pathmorphing}
\usepackage[most]{tcolorbox}
\usepackage[font=small,labelfont=bf]{caption}
\usepackage{todonotes}
\usepackage{hyperref}
\hypersetup{colorlinks=true,linkcolor=blue,citecolor=blue,urlcolor=blue}
\usepackage[capitalize,noabbrev]{cleveref}

\definecolor{chordgray}{RGB}{205,209,215}
\definecolor{editred}{RGB}{205,45,45}
\definecolor{pathorange}{RGB}{230,130,20}

\newcommand{\bigoh}{\mathcal{O}}
\newcommand{\Rbb}{\mathbb{R}}
\newcommand{\dist}{\operatorname{dist}}
\newcommand{\sgn}{\operatorname{sign}}

\newcommand{\Pcal}{\mathcal{P}}
\newcommand{\Qcal}{\mathcal{Q}}
\newcommand{\Ccal}{\mathcal{C}}

\newcommand{\Wone}{\textsf{W[1]}}
\newcommand{\probSPtoEuc}{\textsc{Shortest-Path Euclidean Reweighting}}

\DeclareMathOperator{\poly}{poly}

\newtheorem{theorem}{Theorem}
\newtheorem{lemma}{Lemma}
\newtheorem{proposition}{Proposition}
\newtheorem{corollary}{Corollary}

\title{How to Make Shortest-Path Metrics Euclidean by Reweighting a Few Edges}

\author{Matthias Bentert \and Fedor V. Fomin \and Petr A. Golovach \and M. S. Ramanujan}
\date{}

\begin{document}
\maketitle

\begin{abstract}
Shortest-path metrics infer global distances from local measurements along a graph, while Euclidean metrics describe distances between points in space.
Many applications such as molecular structure determination and sensor localization motivate the question of whether the shortest-path distances inferred from local measurements admit a Euclidean realization.
When it is not possible to embed a given shortest-path metric in Euclidean metric space, can changing a few edge weights fix the issue?
We formalize this question as \probSPtoEuc: given a connected graph~$G$ with positive edge weights and integers~$k$ and~$d$, can changing at most $k$ edge weights make the shortest-path metric of~$G$ isometrically embeddable into~$\Rbb^d$?

As our main result, we show that \probSPtoEuc{} is polynomial-time
solvable for every constant~$k$, independently of~$d$.
Our algorithm runs in $2^{\bigoh(k^2)}n^{\bigoh(k)}$ arithmetic operations.
On rational input, we can also return exact replacement weights.
The dependence on~$k$ in the exponent of~$n$ is presumably necessary,
as we also show that the problem is \Wone-hard parameterized by~$k$ for any~$d \geq 1$.

More generally, our algorithmic techniques enable us to show that for every fixed metric class~$\Ccal$ characterized by
polynomial tests on distances within bounded-size subsets, we obtain
a bound of~$2^{O_{\Ccal}(k^2)}n^{O_{\Ccal}(k)}$ arithmetic operations.
This includes tree metrics, metrics with Gromov hyperbolicity at most
a fixed rational~$\delta \geq 0$, and metrics that embed isometrically
in the Manhattan plane.
\end{abstract}

\medskip

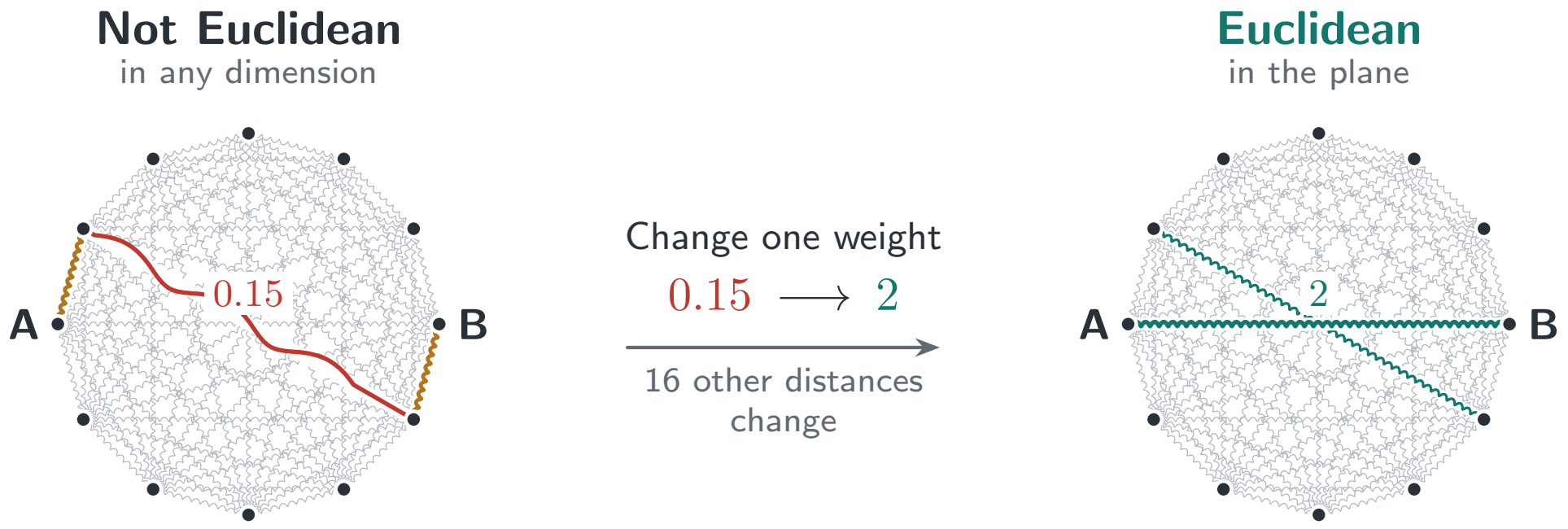
\begin{center}
\begingroup%
\definecolor{figInk}{RGB}{42,48,55}%
\definecolor{figGray}{RGB}{178,184,190}%
\definecolor{figMuted}{RGB}{98,105,112}%
\definecolor{figRed}{RGB}{190,56,47}%
\definecolor{figGold}{RGB}{177,117,27}%
\definecolor{figTeal}{RGB}{18,119,109}%
\begin{tikzpicture}[x=1cm,y=1cm,
  font=\sffamily\fontsize{9}{10.5}\selectfont,
  text=figInk, line cap=round, line join=round,
  figRest/.style={decorate,decoration={coil,aspect=.4,
    amplitude=.5pt,segment length=2.1pt,pre length=1.2pt,post length=1.2pt}},
  figShort/.style={decorate,decoration={coil,aspect=.4,
    amplitude=1.5pt,segment length=6.3pt,pre length=1.2pt,post length=1.2pt}},
  figWide/.style={decorate,decoration={coil,aspect=.4,
    amplitude=2pt,segment length=27pt,pre length=3pt,post length=3pt}},
  figNote/.style={font=\sffamily\fontsize{8}{9.3}\selectfont,text=figMuted},
  figTitle/.style={font=\sffamily\bfseries\fontsize{10.5}{12}\selectfont}]

\node[figTitle] at (2.55,6.36) {Not Euclidean};
\node[figNote] at (2.55,5.98) {in any dimension};
\node[figTitle,text=figTeal] at (11.25,6.36) {Euclidean};
\node[figNote] at (11.25,5.98) {in the plane};

\foreach \figSide/\figCenter in {L/2.55,R/11.25} {
  \foreach \figI in {0,...,11} {
    \coordinate (fig\figSide\figI) at
      ({\figCenter+1.55*cos(150+30*\figI)},
       {3.95+1.55*sin(150+30*\figI)});
  }
  \foreach \figI in {0,...,10} {
    \pgfmathtruncatemacro{\figNext}{\figI+1}
    \foreach \figJ in {\figNext,...,11} {
      \draw[figGray,line width=.24pt,figRest]
        (fig\figSide\figI)--(fig\figSide\figJ);
    }
  }
}
\draw[white,line width=3pt] (figL1)--(figL0)--(figL6)--(figL7);
\draw[figGold,line width=.9pt,figRest] (figL1)--(figL0);
\draw[figRed,line width=1pt,figWide] (figL0)--(figL6);
\draw[figGold,line width=.9pt,figRest] (figL6)--(figL7);
\node[fill=white,inner sep=2pt,text=figRed] at (2.55,4.20) {$0.15$};
\draw[white,line width=2.4pt] (figR0)--(figR6);
\draw[figTeal,line width=.6pt,figRest] (figR0)--(figR6);
\draw[white,line width=3pt] (figR1)--(figR7);
\draw[figTeal,line width=1pt,figRest] (figR1)--(figR7);
\node[fill=white,inner sep=2pt,text=figTeal] at (11.25,4.20) {$2$};

\foreach \figSide in {L,R} {
  \foreach \figI in {0,...,11} {
    \fill[white] (fig\figSide\figI) circle (2.3pt);
    \fill[figInk] (fig\figSide\figI) circle (1.45pt);
  }
  \node[anchor=east,inner sep=4pt,font=\sffamily\bfseries]
    at (fig\figSide1) {A};
  \node[anchor=west,inner sep=4pt,font=\sffamily\bfseries]
    at (fig\figSide7) {B};
}

\node[align=center] at (6.90,4.64) {Change one weight};
\node[font=\sffamily\fontsize{11}{13}\selectfont] at (6.90,4.19)
  {\textcolor{figRed}{$0.15$}\enspace$\longrightarrow$\enspace\textcolor{figTeal}{$2$}};
\draw[figMuted,line width=.8pt,-{Stealth[length=2mm]}]
  (5.63,3.76)--(8.17,3.76);
\node[align=center,figNote] at (6.90,3.31) {16 other distances\\change};
\end{tikzpicture}%
\endgroup%
\end{center}

\medskip

\section{Introduction}\label{sec:intro}

Shortest-path metrics and Euclidean metrics encode distance in two
different ways. A weighted graph gives an implicit metric, obtained by
taking shortest paths. A point set in Euclidean space gives an explicit
coordinate representation. Many applications pass through both
representations: local measurements define a graph, shortest paths infer
global distances, and the final object is expected to admit a Euclidean
realization.

A motivating example comes from molecular structure determination.
Nuclear Magnetic Resonance (NMR) spectroscopy, for which Kurt
Wüthrich was awarded the 2002 Nobel Prize in Chemistry, provides
approximate upper bounds on distances between pairs of nearby hydrogen
atoms via the Nuclear Overhauser Effect (NOE), typically up to
$5$--$6$\,\AA~\cite{Wuthrich1989}. When only a small fraction of the pairwise distances are measured, the
graph formulation of Dress and Havel~\cite{DressHavel1988} propagates the
bounds to all pairs. The graph $G$ has one vertex per atom and takes the
measured NOE bounds as edge weights. The tightest upper bounds derivable
from triangle inequalities are precisely the all-pairs shortest-path
distances in $G$. These bounds are used as input to distance-geometry and
optimization procedures for constructing three-dimensional molecular
conformations~\cite{CrippenHavel1988}.

Even when the measured bounds are valid, the inferred bounds may be
loose and need not coincide with the exact pairwise distances of any
configuration in $\Rbb^3$. Errors such as misassigned cross-peaks or
imperfect calibration can also propagate through the shortest-path
computation. This motivates asking whether changing a few edge weights
can make the resulting shortest-path metric Euclidean.

Related geometric problems arise in sensor network
localization~\cite{AspnesEren2006} and manifold learning. For example,
Isomap uses shortest paths in a neighborhood graph to estimate
geodesic distances~\cite{Tenenbaum2000}.
This gives rise to the following formal problem definition, which is the basis of our work.
Therein, we denote by $\dist_{G,w}$ the shortest-path metric for a connected graph~$G$ with positive edge weights~$w$ and an embedding is \emph{isometric} if it preserves every pairwise distance.

\smallskip
\begin{tcolorbox}[
breakable,
enhanced,
colback=white,
colbacktitle=white,
colframe=black,
coltitle=black,
boxrule=0.5pt,
arc=3pt,
left=2mm, right=2mm, top=1.5mm, bottom=1.5mm,
]
\noindent\textbf{\probSPtoEuc{}}

\smallskip
\noindent\textit{Input.} A connected graph $G$, a positive edge-weight
function $w\colon E(G)\to\Rbb_{>0}$, and integers $k\ge0$ and $d\ge 1$.

\smallskip
\noindent\textit{Question.} Is there a set $F\subseteq E(G)$ with
$|F|\le k$ and a positive edge-weight function~${w'\colon E(G)\to\Rbb_{>0}}$ such that $w'(e)=w(e)$ for each~$e\in E(G)\setminus F$ and the shortest-path metric~$\dist_{G,w'}$ embeds isometrically into~$\Rbb^d$?
\end{tcolorbox}

\begin{figure}[t]
\centering
\begin{tikzpicture}[scale=.95]
\def\Rad{1.8}
\draw[gray!30] (0,0) circle (\Rad);
\foreach \k in {0,...,15}{\coordinate (P\k) at ({90-22.5*\k}:\Rad);}
\draw[gray!60,dashed,line width=.8pt] (P1)--(P10);
\draw[blue!70!black,very thick] (P1)--(P0);
\draw[blue!70!black,very thick] (P8)--(P10);
\draw[red!75!black,very thick] (P0)--node[left,pos=.3,font=\small]{$\delta$}(P8);
\foreach \k in {0,...,15}{\fill (P\k) circle (1.5pt);}
\foreach \k in {0,1,8,10}{\node[font=\small] at ({90-22.5*\k}:{\Rad+.38}) {$v_{\k}$};}
\end{tikzpicture}
\caption{The construction for $n=16$. Every edge of the complete graph
has its Euclidean length as its weight, except for the diameter $v_0v_8$
(red), which has weight $\delta$. Only a few edges are shown. The
shortest path from $v_1$ to $v_{10}$ (shown in blue and red) uses the
diameter and is shorter than the direct edge (dashed). The same holds
for $\Omega(n^2)$ pairs, so restoring the Euclidean weight of the
diameter changes the shortest-path distance for each of these pairs.}
\label{fig:cascade}
\end{figure}
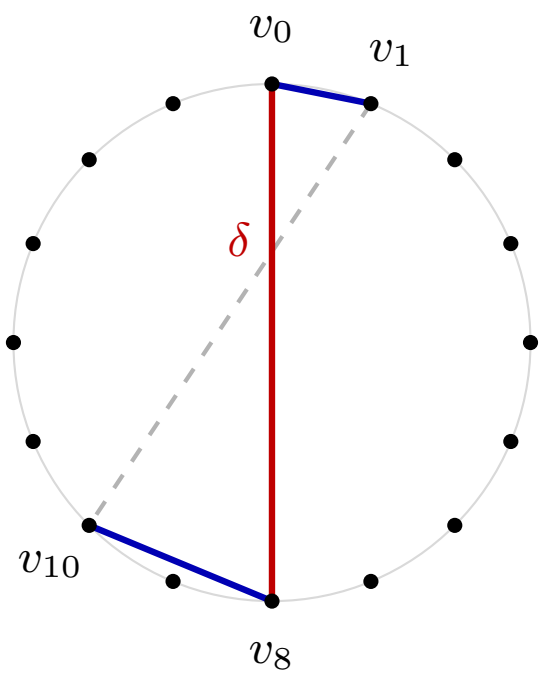

We note that in \probSPtoEuc{}, \emph{a single edge-weight edit can change many
entries of the shortest-path distance matrix}. 
The following example shows that the number of changed distances can be quadratic.
See also \Cref{fig:cascade} for an illustration.
Let $n\ge8$ be even, and place~$v_0,\ldots,v_{n-1}$ in this order at the
vertices of a regular $n$-gon inscribed in a circle of radius $\rho$ in
the plane. In the complete graph on these vertices, assign each edge
its Euclidean length, except for the diameter $e^*=v_0v_{n/2}$, which
receives weight $0<\delta<\rho/4$. For $0\le i,j\le n/8$, the path from~$v_i$
through~$v_0$ and~$v_{n/2}$ to~$v_{n/2+j}$ has length at most
$4\rho\sin(\pi/8)+\delta<1.79\rho$. The Euclidean distance between $v_i$
and $v_{n/2+j}$ is
$2\rho\cos(\pi(j-i)/n)\ge2\rho\cos(\pi/8)>1.84\rho$.
Thus, for $\Omega(n^2)$ pairs, the shortest-path distance is smaller
than the Euclidean distance. Changing the weight of $e^*$ to $2\rho$
makes each edge a shortest path between its endpoints by the triangle
inequality. The shortest-path metric then coincides with the Euclidean
metric on the vertices of the polygon. One edge-weight edit therefore
changes $\Omega(n^2)$ distances.

\subsection{Our results }
When all edge weights may be changed, feasibility has a purely
combinatorial characterization: for every $d\ge1$, a connected graph
admits positive edge weights whose shortest-path metric embeds
isometrically into $\Rbb^d$ if and only if it contains a Hamiltonian
path (\Cref{prop:unrestricted-hamiltonian}).

We next ask whether such a reweighting can be achieved by changing
at most $k$ edge weights. Our main algorithmic result gives a
polynomial-time algorithm for \probSPtoEuc{} for every fixed $k$,
even when the target dimension $d$ is part of the input.

\begin{restatable}{theorem}{euclideanTheorem}\label{thm:main}
\probSPtoEuc{} can be solved in $2^{\bigoh(k^2)}n^{\bigoh(k)}$ arithmetic operations, where $n=|V(G)|$.
For rational input weights, it can be solved with~$2^{\bigoh(k^2)}n^{\bigoh(k)}\poly(L)$ bit operations, and the algorithm can return exact weights for the edited edges on yes-instances.
\end{restatable}

The bit bound is for binary rational input weights and~$L\ge2$ denotes the total binary encoding length of the numerical input, including the budget $k$ and the target dimension. 
In both models, the new weight assigned to an edited edge may be any positive real number.
We complement our algorithm with a parameterized hardness result,
which rules out fixed-parameter tractability in $k$ unless
$\mathsf{FPT}=\Wone$.

\begin{restatable}{theorem}{euclideanHardnessTheorem}\label{thm:euclidean-hardness}
\probSPtoEuc{} is $\Wone$-hard parameterized by $k$ for any~$d \geq 1$, even when all input weights are positive integers encoded in unary, a Hamiltonian path is given as part of the input, and the input graph is restricted to be triangle-free.
\end{restatable}

The algorithm in \Cref{thm:main} separates two tasks. After fixing the edges
that may be edited, we divide the space of possible new weights into
regions on which every shortest-path distance is an affine function of
those weights. On each region, we then express Euclidean embeddability
by polynomial comparisons. The first step depends only on shortest
paths, so the same approach applies to other classes of metrics whose
membership can be tested by polynomial comparisons on subsets of
bounded size.
Let $\Ccal$ be a class of finite metrics. We introduce the following
problem.

\begin{tcolorbox}[enhanced,colback=white,colframe=black,boxrule=0.5pt,
  arc=2pt,left=2mm,right=2mm,top=1.5mm,bottom=1.5mm]
\noindent\textbf{Shortest-path reweighting into $\Ccal$}

\smallskip
\noindent\emph{Input.} A connected graph $G=(V,E)$, a positive edge-weight
function $w\colon E\to\Rbb_{>0}$, and an integer $k\ge0$.

\smallskip
\noindent\emph{Question.} Is there a set $F\subseteq E$ with $|F|\le k$
and a positive edge-weight function $w'\colon E\to\Rbb_{>0}$ such that
$w'(e)=w(e)$ for every $e\in E\setminus F$ and $\dist_{G,w'}\in\Ccal$?
\end{tcolorbox}

Our last main result is then the following.

\begin{restatable}{theorem}{localTheorem}\label{thm:local}
For every fixed class $\Ccal$ of finite metrics with a local polynomial
description, shortest-path reweighting into $\Ccal$ can be solved using
$2^{O_{\Ccal}(k^2)}n^{O_{\Ccal}(k)}$ arithmetic operations. On rational
input, it can be solved using
$2^{O_{\Ccal}(k^2)}n^{O_{\Ccal}(k)}L^{O_{\Ccal}(1)}$ bit operations,
and on yes-instances the algorithm can return exact real algebraic
weights for the edited edges.
\end{restatable}

Therein, a \emph{local polynomial description} of $\Ccal$ consists of an integer
$r\ge1$ and fixed quantifier-free Boolean formulas~$\Phi_1,\ldots,\Phi_r$ with the following properties.
\begin{enumerate}
\item A finite metric $D$ belongs to $\Ccal$ if and only if
      $D|_U\in\Ccal$ for every nonempty subset $U$ with~${|U|\le r}$.
\item For each $1\le t\le r$, the formula $\Phi_t$ tests membership in
      $\Ccal$ on a $t$-point metric using polynomial comparisons in its
      $\binom{t}{2}$ pairwise distances.
\end{enumerate}
Constants indexed by $\Ccal$ may depend on $r$ and on the sizes, degrees, and coefficients of the formulas.
A \emph{polynomial comparison} is an equality or inequality comparing a
polynomial with rational coefficients to zero. A quantifier-free Boolean
formula combines such comparisons using conjunction, disjunction, and
negation. For a metric $D$, write $D|_U$ for its restriction to the
subset $U$ of its point set.

Examples of metrics handled by \Cref{thm:local} include ultrametrics, tree metrics, metrics with Gromov hyperbolicity at most a fixed rational $\delta\ge0$, and metrics that embed isometrically in the Manhattan plane.
For fixed~$d$, Menger’s characterization gives a polynomial description of Euclidean embeddability using subsets of at most~$d+3$ points.
\cref{thm:main} gives an exponent independent of~$d$.

\subsection{Prior work on metric embeddings and editing}

\subparagraph{Metric embeddings.}
A large body of work embeds finite metrics, and in particular
shortest-path metrics of graphs, into Euclidean or $\ell_p$ spaces with
small distortion~\cite{Bourgain1985,linial1995geometry,AbrahamFGN18}.
The survey of Indyk, Matou\v{s}ek, and Sidiropoulos~\cite{indyk20178}
gives an overview. In that setting the input metric is fixed and distances
may be distorted. We require an isometric embedding and allow a few edge weights
to change.

\subparagraph{Distance geometry.}
Exact Euclidean realization is characterized by the Cayley--Menger and
Gram-matrix
criteria~\cite{Cayley1841,Menger1928,Schoenberg1935,YoungHouseholder1938,Blumenthal1970}.
Bowers and Bowers~\cite{BowersBowers2017} give a complete proof of
Menger's bounded-subset theorem. Distance geometry often starts from
partial distance information~\cite{LibertiLavorMaculanMucherino2014,CrippenHavel1988}.
For example, network localization uses specified neighbor distances and
beacon positions~\cite{AspnesEren2006}. In our problem, the input graph
and weights determine the distances between all vertex pairs by shortest
paths.

\subparagraph{Metric editing and completion.}
Fan, Gilbert, Raichel, Sonthalia, and Van Buskirk~\cite{FanGRSB20} also edit weights on a possibly incomplete graph. Their goal is to change weights so that no cycle has an edge longer
than the rest of the cycle combined. Equivalently, each final edge must
realize the shortest-path distance between its endpoints. Here an edge
may remain longer than a path between its endpoints. It is the induced
metric on all vertex pairs that must belong to the target class.

Sparse metric editing asks whether changing at most $k$ pairwise
distances can enforce the triangle inequality~\cite{FanRB18,gilbert2017sparse}.
Related problems fit a given distance matrix by a metric from a prescribed
class~\cite{BergerKL99,SidiropoulosWW17,CohenAddadTreeFitting}.
For the objective of minimizing the number of changed entries, constant-factor
approximations are known for fitting ultrametrics~\cite{charikar2024improved,AnKLL25}
and tree metrics~\cite{Kipouridis23}. Constant-factor approximations also hold
for both classes when lower bounds are imposed on the output distances~\cite{DasKS26}. 
Euclidean distance-matrix completion instead fills missing distances
consistently with fixed
ones~\cite{Bakonyi1995,Alfakih1999,Laurent2001,FominGRS26EDMC}. Bentert, Fomin, Golovach, Ramanujan, and Saurabh~\cite{BentertFGRS25} use algebraic feasibility for
Euclidean distance-entry editing with possible outlier deletions, and
Fomin, Golovach, Ramanujan, and Saurabh~\cite{FominGRS26EDMC} use it to test whether the fill-in entries in distance-matrix completion
can be assigned feasible values. In our setting the
unknowns are edge weights. Expressing all shortest-path distances in
those weights on a common region is what lets us apply the same tools.

\subsection{Our methods}
\label{sec:overview}
We next give a high-level overview of how we prove our main results.

\paragraph*{Hamiltonian paths and unrestricted reweighting} We start with the structural observation that embeddability in~$\Rbb^d$ under an unbounded number of weight modifications is equivalent to containing a Hamiltonian path independent of~$d$.
It is based on the observation that the vertices of any shortest path must be embedded in path order on a straight line segment between its endpoints.
If the graph contains a Hamiltonian path, then we can embed all vertices on a line along that Hamiltonian path and change the weights to match the distance between their endpoints.
In the other direction, if the shortest-path metric has a Euclidean realization in some dimension~$d \geq 1$ after changing the weights of edges, then we can sort all vertices by the lexicographic order of their coordinates.
Between each two consecutive vertices in this order, there must be an edge as the shortest path between them must be embedded on a straight line and no vertex can be embedded on that line between the two vertices without being placed between them in the lexicographic order.
This then shows that a Hamiltonian path exists in the input graph.

\paragraph*{Hardness on the line and in higher dimensions} We next turn to \Cref{thm:euclidean-hardness} which shows that the dependence on $k$ in the exponent of~$n$ cannot be removed unless~$\mathsf{FPT}=\Wone$.
Here, we reduce from \textsc{Multicolored Independent Set}, where the goal is to choose one vertex from each of~$\ell$ color classes with no edge between any two chosen vertices.
Each color class is represented by a unit-weight cycle with two additional vertices serving as anchors.
Moreover, we add heavy edges between vertices in different color cycles based on the edges of the input graph. 
The budget is set to~$k=3\ell$ and we enforce that the weight of exactly one edge of each cycle and exactly one edge between each cycle and each of its two anchors must be modified.
The new weight of the modified edge inside the cycle must then be made large enough to allow the complete cycle to be placed between the two endpoints.
The choice of which edge weight is modified encodes the selected vertex for each color.
The anchors and the paths between them force the cycles to appear as consecutive blocks in color order.
If two selected vertices were adjacent in the input, the edge representing that adjacency would be shorter than the forced distance between its endpoints.
This allows us to show that a solution for~$d=1$ exists if and only if the input graph contains a multicolored independent set.
The constructed graph is also triangle-free.
We then show that any triangle-free graph can be embedded isometrically in~$\Rbb^d$ if and only if it can be embedded isometrically on a line.
This holds as otherwise, a triangle of minimum positive area would force edges between all its endpoints, contradicting triangle-freeness.
This shows \Wone-hardness for any target dimension.

\paragraph*{Polynomial-time algorithms for a fixed edit budget}
Finally, we come to \Cref{thm:main,thm:local}.
Both algorithms start identically.
We first guess a set~$F=\{e_1,e_2,\ldots,e_k\}$ of at most~$k$ edges whose weights may be modified.
We replace their weights by variables~$x_1,x_2,\ldots,x_k$.
For each pair~$u,v$ of vertices and each~$S\subseteq[k]$, let~$c_{uv,S}$ be the minimum length of a $u$-$v$-walk that traverses each~$e_i$ with~$i\in S$ exactly once and uses no other edge of~$F$, where we ignore the weights~$x_i$ with~$i \in S$.
If no such walk exists, we set~$c_{uv,S} = \infty$.
Moreover, the order in which edges in~$S$ are used and their orientation might matter, but there are only a limited number (in terms of~$k$) possible choices and we can iterate over all of them.
Finally, if after fixing values for the variables~$x_i$ a shortest~$u$-$v$-path uses precisely the edges in~$S$, then the distance between~$u$ and~$v$ is~$c_{uv,S}$ plus the sum of the fixed weights~$x_i$ with~$i \in S$.
When the~$x_i$ are considered variables, the distance between~$u$ and~$v$ is then~$D_{uv}(x) = \min_{S \subseteq [k]} \left( c_{uv,S} + \sum_{i \in S} x_i\right)$.
Note that~$D(x)$ is an~$n \times n$~matrix where each entry is piecewise a polynomial of degree at most one in~$x$.
In particular, each entry of~$D(x)$ is a polynomial of degree at most one in~$x$ on any region where one minimizing term is fixed for every vertex pair.

Before we present the next step of both algorithms, let us imagine a case where an oracle told us for each pair of vertices, which of the~$k$ chosen edges are part of a shortest path between them.
Then, the task would boil down to finding weights for the~$k$ chosen edges such that the shortest paths are indeed through the specified edges and the resulting matrix~$D(x)$ belongs to the target class.
Without this oracle, there is an additional complication: it is not clear which pairs of vertices impose restrictions on the weights of which of the~$k$ chosen edges.
The next goal is to simulate such an oracle.
The basic observation is that for a given weight vector~$x$, a term indexed by~$S$ with $c_{uv,S}<\infty$ attains the minimum for a pair~$u,v$ if and only if~$c_{uv,S}+\sum_{i \in S}x_i \leq c_{uv,T}+\sum_{i \in T}x_i$ for each~$T \subseteq [k]$ with $T\ne S$ and $c_{uv,T}<\infty$.
In principle, there are at most~$n^2 4^k$ such comparisons possible and each comparison can return one of three possible outcomes:~$<, =$, or~$>$.
Hence, guessing the outcomes of these comparisons naively gives at most~$3^{n^24^k}$ guesses to verify.
Fortunately, we can observe that each comparison tests the sign of a polynomial of degree one in~$x$.
Hence, each such comparison can be seen as a plane that splits space of weights~$\Rbb^k$ in three sets: points above the plane, points below the plane, and points on the plane.
Now, we can make use of the following result by Basu, Pollack, and Roy~\cite{BasuPR06}.
\begin{lemma}[{\cite{BasuPR06}}]\label{fact:bpr}
Let $\Pcal$ be a set of $s$ polynomials of degree at most~$\Delta\ge1$ in $k\geq 1$
variables. The realizable sign conditions on~$\Pcal$ can be computed using
$(s\Delta+2)^{\bigoh(k)}$ arithmetic operations.
If the coefficients are integers of bit length at most~$\tau$, the computation uses $\poly(\tau)(s\Delta+2)^{\bigoh(k)}$ bit operations.
If the coefficients are rational, one can also return an exact real point in each realizable sign condition in the same time.
\end{lemma}
Applying this result with~$\Delta=1$ allows us to enumerate the realizable sign conditions in~$2^{\bigoh(k^2)}n^{\bigoh(k)}$ arithmetic operations.
For each such condition, we choose one minimizing term for every vertex pair, breaking ties arbitrarily.
Let~$R$ be the set of positive weight vectors for which all chosen terms attain their respective minima.
It remains to test whether some~$x\in R$ yields a metric in the target class.

\paragraph*{Reweighting into classes with local polynomial descriptions}
Here, the proofs of \Cref{thm:main,thm:local} diverge.
For \Cref{thm:local}, the requirement that the target class has a local polynomial description simplifies the proof significantly.
For a fixed class characterized by polynomial tests on subsets of at most $r$ points, we can substitute the distances~$D_{uv}(x)$ into all those tests and require them to hold
simultaneously for the same weight vector.
There are $\bigoh_r(n^r)$ subsets, and affine substitution does not increase the degrees of the fixed tests.
Whether such a vector exists can then again be checked using~\Cref{fact:bpr}.

\paragraph*{Euclidean reweighting with the dimension as input}
For \Cref{thm:main}, we could do the same and test each set of at most~$d+3$ points, where the target metric space is Euclidean~$\Rbb^d$.
This would result in a running time of the form~$f(k,d)n^{\bigoh(kd)}$ for any~$k,d \geq 1$.
Note that \Cref{thm:main} promises something much stronger: a running time of the form~$2^{\bigoh(k^2)}n^{\bigoh(k)}$.
Instead of testing all subsets of at most~$d+3$ points, we test the entire distance matrix at once using the Gram criterion as explained next.
On a fixed region, this gives polynomial conditions in the~\(k\) replacement weights, without introducing coordinates for the vertices.
The Gram criterion states that a symmetric matrix~$D$ with nonnegative entries and zeros on its main diagonal can be embedded isometrically in~$\Rbb^d$ if and only if a certain matrix~$B$ is positive semidefinite and of rank at most~$d$~\cite{Schoenberg1935,YoungHouseholder1938}.
The entries of this~$(n-1)\times(n-1)$~matrix~$B$ are defined by~$B_{uv} = \tfrac12\bigl(D_{uo}^2+D_{vo}^2-D_{uv}^2\bigr)$ where it is assumed that some vertex~$o$ is positioned at the origin and~$u,v \in V \setminus \{o\}$.
Note that the entries of~$B(x)$ are defined via squares of entries in~$D(x)$ and are therefore polynomials of degree at most~$2$ in~$x$.
To decide whether~$B(x)$ is positive semidefinite and of rank at most~$d$, we use the characteristic polynomial~$\det(tI+B(x)) = t^{n-1}+c_1(x)t^{n-2}+\cdots+c_{n-1}(x)$ of~$-B(x)$.
We can then use the well known fact that a real symmetric matrix (which~$B(x)$ is) is positive semidefinite and of rank at most~$d$ if and only if all coefficients~$c_j(x)$ in the characteristic polynomial of~$-B(x)$ are nonnegative and (if~$d<n-1$) $c_{d+1}(x)=0$.
We show that each~$c_j(x)$ has degree at most~$2j$, and we compute all these polynomials with~$n^{O(k)}$ arithmetic operations using Newton’s identities.
To conclude the proof overview, we test whether these two conditions hold for some assignment of the variables in~$x$ in the specified region~$R$ using \Cref{fact:bpr}.

\section{Preliminaries}\label{sec:prelim}

All graphs considered in this work are finite, simple, and undirected.
We denote the number of vertices in a graph~$G$ by~$n$ and the number of edges by~$m$.
For an integer~$t$, let~$[t]=\{1,\ldots,t\}$.
For~$x\in\Rbb^k$ and $S\subseteq[k]$, we write~$x(S)=\sum_{i\in S}x_i$.
For a graph~$H$ with positive edge weights, we denote the length of a shortest $a$-$b$-path in~$H$ by $\dist_H(a,b)$.

\subparagraph{Matrices and sign conditions.}
A symmetric matrix $B$ is \emph{positive semidefinite} if~${\alpha^{\mathsf T}B\alpha\geq 0}$ for each real vector $\alpha$.
Equivalently, all of its eigenvalues are nonnegative.

Let~$\Pcal$ be a finite set of real polynomials in $x=(x_1,\ldots,x_k)$.
A \emph{sign condition} on~$\Pcal$ assigns to every $P\in\Pcal$ a value
$\sigma(P)\in\{-1,0,1\}$. A point $x\in\Rbb^k$ \emph{realizes} $\sigma$ if
$\sgn (P(x))=\sigma(P)$ for each~$P\in\Pcal$, and $\sigma$ is
\emph{realizable} if some point realizes it.

\subparagraph{Computational models.}
In the arithmetic model, an input weight or numerical parameter is
available as a single number. Addition, subtraction, multiplication,
division by a nonzero number, and comparison are exact unit-cost
operations. We also allow the usual indexing and traversal of the
explicit graph. In the bit model, rational inputs are encoded by binary
integer numerators and positive denominators, and integer inputs such
as $k$ and $d$ are encoded in binary. We count bit operations on these
encodings, including the cost of reading them and writing the output. We take
$L$ to be at least two. All bit bounds below assume rational numerical
data. Feasible solutions may use irrational weights.
The graph can be encoded by an edge list with vertex indices of
$\bigoh(\log n)$ bits, so its encoding length is polynomial in $n$.
The formulas for a fixed target class are part of the algorithm.
Claims about complexity classes and reductions use finite binary inputs.

\section{The unrestricted editing case}

Before turning to the parameterized algorithm, we study what happens when all edge weights are allowed to change.
The answer for Euclidean spaces is surprisingly combinatorial and says that the existence of a Hamiltonian path is equivalent to embeddability of the shortest-path metric in any dimension.

\begin{proposition}\label{prop:unrestricted-hamiltonian}
Let $G$ be a connected graph. The following are equivalent:
\begin{enumerate}
\item $G$ contains a Hamiltonian path;
\item there exists an integer~$d \geq 1$ and a positive edge-weight function~$w\colon E(G)\to\Rbb_{>0}$ such that $\dist_{G,w}$ embeds isometrically into $\Rbb^d$;
\item for all integers~$d \geq 1$, there exists a positive edge-weight function $w\colon E(G)\to\Rbb_{>0}$ such that $\dist_{G,w}$ embeds isometrically
into $\Rbb^d$.
\end{enumerate}
\end{proposition}

\begin{proof}
It is trivial to see that (3) implies (2).
We next show that (2) implies (1). 
Assume that we can assign positive weights to edges in~$G$ such that the resulting weighted graph gives a shortest-path metric that can be embedded into~$\mathbb{R}^d$ for some~$d \geq 1$.
Let~$p_i=(c^1_i,c^2_i,\ldots,c^d_i)$ be the point in~$\mathbb{R}^d$ onto which~$v_i$ is embedded for each~$v_i \in V$. 
We next describe an ordering of the vertices.
For two vertices~$v_i$ and~$v_j$, let~$d_{i,j}$ be the smallest integer such that~$c^{d_{i,j}}_i \neq c^{d_{i,j}}_j$.
Note that since all vertices are embedded onto distinct points in~$\mathbb{R}^d$ (recall that all pairwise distances between distinct vertices are positive, as $G$ is connected with positive edge weights), the integer~$d_{i,j} \leq d$ exists for all~$i \neq j \in [n]$.
We say that~$v_i$ comes before~$v_j$ in our ordering if~$c^{d_{i,j}}_i < c^{d_{i,j}}_j$.
Note that this describes a total ordering of all vertices.
We will show that this ordering describes a Hamiltonian path.
To this end, consider two vertices~$v_i,v_j$ such that no vertex is between the two in our ordering and~$v_i$ comes before~$v_j$ in the ordering.
We will show that the edge~$v_iv_j$ exists in~$G$.
Assume for a contradiction that this is not the case.
Then, there must be a path between~$v_i$ and~$v_j$ whose length is exactly
$\|p_i-p_j\|_2$.
Note that all internal vertices of that path are embedded in the open line segment between~$p_i$ and~$p_j$.
Thus, for each such vertex~$v_\ell$ of this path and each~$d' < d_{i,j}$, it holds that~$c^{d'}_{i} = c^{d'}_{\ell} = c^{d'}_j$.
Moreover, it holds that~$c^{d_{i,j}}_{i} < c^{d_{i,j}}_{\ell} < c^{d_{i,j}}_j$.
Thus, $v_{\ell}$ appears between~$v_i$ and~$v_j$ in our ordering, a contradiction.
This proves that our ordering describes a Hamiltonian path.

To conclude the proof, we show that (1) implies (3). 
Assume that~$G$ contains a Hamiltonian path~$\mathcal{P}$.
We rename the vertices such that the Hamiltonian path becomes~$\mathcal{P}=(v_1,v_2,v_3,\ldots,v_n)$.
We then set the weight of any edge~$v_iv_j$ to~$|j-i|$.
Note that since~$\mathcal{P}$ is a Hamiltonian path, there exists an edge~$v_iv_{i+1}$ for each~$i \in [n-1]$.
Moreover by construction, the distance between any two vertices~$v_i$ and~$v_j$ with~$j \geq i$ is exactly~$j-i$.
Hence, we can embed them isometrically on a line by placing each vertex~$v_i$ at position~$i$.
We then also have the additional property that any edge is a shortest path between its two endpoints.
Thus we obtain an isometric embedding into $\mathbb{R}^1$ and since this line also exists in~$\mathbb{R}^d$ for any~$d \geq 1$, this concludes the proof.
\end{proof}

\section{Shortest-path distances as functions of the edited weights}\label{sec:machinery}

We now present our main results.
In this section, we lay the groundwork that enables both \cref{thm:main,thm:local}.
In \cref{sec:local}, we then show \cref{thm:local} as a direct consequence and give some examples of metrics to which the theorem applies.
In \cref{sec:euclid}, we then show how to prove \cref{thm:main}.
Throughout the exposition, we assume that~$k \geq 1$ as the problem with~$k=0$ is trivial.

\subsection{Distances as minima of affine functions}\label{sec:distances}

Fix a set $F=\{e_1,\ldots,e_k\}$ of $k$ edges of the connected
graph~$G$. For $x\in\Rbb^k_{>0}$, let $G_x$ be the
weighted graph in which $e_i$ has weight $x_i$ for every $i\in[k]$ and
every other edge~$e$ has weight $w(e)$. For vertices $a$ and~$b$, let
$\dist_{G-F}(a,b)$ be the distance between $a$ and~$b$ in $G-F$ with
weights~$w$, and let $\dist_{G-F}(a,b)=+\infty$ if $a$ and~$b$ lie in different
components of $G-F$. We fix a linear order on~$V$, so that every pair of
distinct vertices is written $u<v$ in exactly one way.

\begin{lemma}\label{lem:affine}
For all vertices $u<v$ and all $S\subseteq[k]$ there is a constant
$c_{uv,S}\in\Rbb_{\ge0}\cup\{+\infty\}$ such that every
$x\in\Rbb^k_{>0}$ satisfies
\begin{equation}\label{eq:affine}
  \dist_{G_x}(u,v)=\min_{S\subseteq[k]}\bigl(c_{uv,S}+x(S)\bigr).
\end{equation}
All these constants can be computed using $2^kk!\poly(n)$
arithmetic operations.
\end{lemma}

\begin{proof}
Fix $u<v$. Consider sequences $((a_1,b_1),\ldots,(a_t,b_t))$ with
$0\le t\le k$ such that $a_1b_1,\ldots,a_tb_t$ are distinct edges of~$F$.
Such a sequence describes the order and the directions in which a $u$--$v$
path might traverse edges of~$F$. Its \emph{cost} is
\[
  \dist_{G-F}(u,a_1)+\sum_{j=1}^{t-1}\dist_{G-F}(b_j,a_{j+1})+\dist_{G-F}(b_t,v),
\]
which is $\dist_{G-F}(u,v)$ when $t=0$. Let $c_{uv,S}$ be the minimum cost of a
sequence whose edges are exactly the edges $e_i$ with $i\in S$. There are
at most $\sum_{t=0}^{k}2^tk!/(k-t)!\le e\,2^kk!$ sequences, and $\dist_{G-F}$
can be computed by any all-pairs shortest-path algorithm. This gives the
bound on the number of operations.

Let $x\in\Rbb^k_{>0}$. If $c_{uv,S}$ is finite, take a sequence of
cost~$c_{uv,S}$ and replace every term $\dist_{G-F}(a,b)$ of its cost by a
shortest $a$-$b$-path in $G-F$. Together with the edges of the sequence,
these paths form a $u$--$v$ walk in~$G_x$ of length $c_{uv,S}+x(S)$.
Hence the left side of~\eqref{eq:affine} is at most the right side.

Conversely, since all weights are positive, some shortest $u$--$v$
path~$P$ in~$G_x$ has no repeated vertex, and so it traverses each edge
of~$F$ at most once. List the edges of~$F$ on~$P$ in the order and the
directions in which $P$ traverses them, and let $S$ be the set of their
indices. The parts of~$P$ before, between, and after these edges avoid~$F$.
Each of them is therefore at least as long as the corresponding term in
the cost of the listed sequence. Hence the length of~$P$ is at least
$c_{uv,S}+x(S)$, and the left side of~\eqref{eq:affine} is at least the
right side.
\end{proof}

The constants depend only on the unchanged weights. Connectivity ensures
that at least one term is finite for every vertex pair.

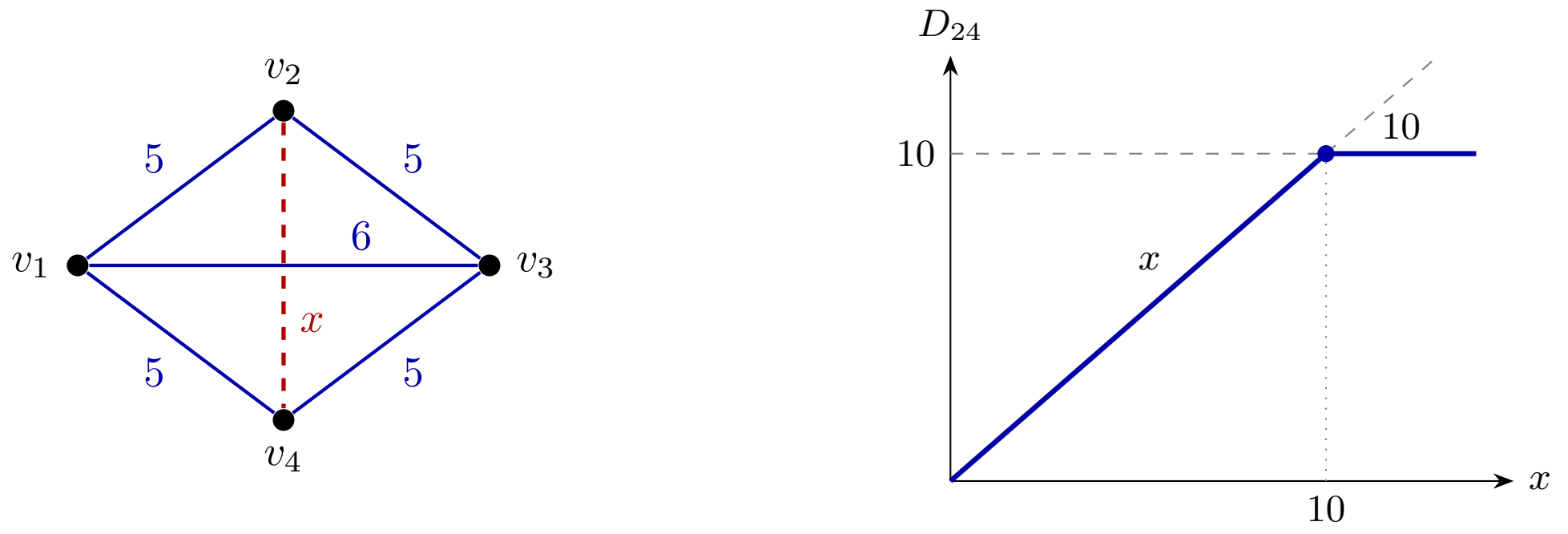
\begin{figure}[t]
\centering
\begin{minipage}[c]{.46\linewidth}
\centering
\begin{tikzpicture}[scale=.85,
 vertex/.style={circle,fill=black,inner sep=1.8pt},
 fixed/.style={line width=.8pt,blue!65!black},
 editable/.style={line width=1pt,red!70!black,dashed}]
\node[vertex,label=left:$v_1$] (A) at (-2,0) {};
\node[vertex,label=above:$v_2$] (B) at (0,1.5) {};
\node[vertex,label=right:$v_3$] (C) at (2,0) {};
\node[vertex,label=below:$v_4$] (D) at (0,-1.5) {};
\draw[fixed] (A)--node[above left] {$5$}(B);
\draw[fixed] (B)--node[above right] {$5$}(C);
\draw[fixed] (C)--node[below right] {$5$}(D);
\draw[fixed] (D)--node[below left] {$5$}(A);
\draw[fixed] (A)--node[above,pos=.7] {$6$}(C);
\draw[editable] (B)--node[right,pos=.7] {$x$}(D);
\end{tikzpicture}
\end{minipage}\hfill
\begin{minipage}[c]{.50\linewidth}
\centering
\begin{tikzpicture}[x=.31cm,y=.27cm,>=Stealth,font=\small]
\draw[->] (0,0)--(15,0) node[right] {$x$};
\draw[->] (0,0)--(0,13) node[above] {$D_{24}$};
\draw[gray,dashed] (0,10)--(14,10);
\draw[gray,dashed] (10,10)--(13,13);
\draw[gray,dotted] (10,0)--(10,10);
\draw[blue!65!black,very thick] (0,0)--(10,10)--(14,10);
\fill[blue!65!black] (10,10) circle (2pt);
\node[below] at (10,0) {$10$};
\node[left] at (0,10) {$10$};
\node[above left] at (6,6) {$x$};
\node[above] at (12,10) {$10$};
\end{tikzpicture}
\end{minipage}
\caption{An edge weight becomes a variable, and a distance becomes a
minimum of affine functions. The breakpoint is determined by competing graph paths,
before any target metric condition is imposed.}
\label{fig:example}
\end{figure}

\subparagraph{One edited edge.}\label{sec:example}
In \Cref{fig:example}, only the edge $v_2v_4$ changes, to weight $x>0$.
Write $D_{ij}$ for the distance between $v_i$ and $v_j$. The five other
pairwise distances remain their edge weights, while
$D_{24}(x)=\min\{x,10\}$. The two affine regions meet at $x=10$.
A target condition may restrict $x$ to a subinterval or a single point
within an affine region. 

\subsection{Enumerating affine regions and testing each for feasibility}\label{sec:template}

By \Cref{lem:affine}, we can divide the space of positive replacement
weights into regions on which all distances are affine. This step is
independent of $\Ccal$. For each region we must then decide whether
some choice of weights in that region yields a metric in $\Ccal$.

\subparagraph{Testing a region for membership in the target class.}
An instance of \emph{affine feasibility for $\Ccal$} consists of a
symmetric $n\times n$ matrix $D(x)$ with zero diagonal and affine
off-diagonal entries in $x\in\Rbb^k$, and a region $R$ described by $s$
linear equalities and weak or strict inequalities. We assume that $D(x)$ is a metric for every $x\in R$. The region may be empty.
The required output is yes exactly when
there exists $x\in R$ such that $D(x)\in \Ccal$. 

We denote by~$T_{\Ccal}(n,k,s)$ a nondecreasing upper bound on the arithmetic
cost of this subroutine, for the given target class. \Cref{sec:local} implements this subroutine with local polynomial tests, and \Cref{sec:euclid} implements it with a global Gram test.

\subparagraph{Enumerating editable edges and distance formulas.}
We may assume $k\le m$, since a larger budget can be replaced by~$m$.
If $n=1$, test the one-point metric directly.
A guessed edge may keep its original weight, so it suffices to consider
sets of exactly $k$ edges. For each such set~$F$, compute the constants
of \Cref{lem:affine}. Form the set $\Qcal$ of affine comparisons
\begin{equation}\label{eq:comparison}
 (c_{uv,S}+x(S))-(c_{uv,T}+x(T))
\end{equation}
over all $u<v$ and all distinct $S,T\subseteq[k]$ for which both constants
are finite. Enumerate the realizable sign conditions on~$\Qcal$ using
\Cref{fact:bpr}. For each such condition~$\sigma$, do the following.
\begin{enumerate}
\item For every $u<v$, choose a set $S_{uv}$ whose affine term is minimal
  according to~$\sigma$, breaking ties arbitrarily. Put
  $\ell_{uv}(x)=c_{uv,S_{uv}}+x(S_{uv})$, and let $D(x)$ have these
  off-diagonal entries and zero diagonal.
\item Define the validity region
  \begin{equation}\label{eq:region}
  R=\left\{x\in\Rbb^k:
  \begin{array}{ll}
    x_i>0 & (i\in[k]),\\
    \ell_{uv}(x)\le c_{uv,T}+x(T)
       & (u<v,\ T\subseteq[k],\ c_{uv,T}<\infty)
  \end{array}\right\}.
  \end{equation}
\item Call the affine-feasibility subroutine for $\Ccal$ on $(D,R)$.
  Accept if it returns yes.
\end{enumerate}
Reject if no call returns yes. The inequalities defining $R$ keep the
chosen term minimal for each pair. Comparisons between other terms need not keep their
original signs, and the weak inequalities
allow several terms to attain the same minimum.
Let~$B(n,k)=2^{\bigoh(k^2)}n^{\bigoh(k)}$ and~$s_0=k+\binom n2 2^k$. 

\begin{theorem}\label{thm:template}
The algorithm correctly decides reweighting into $\Ccal$ in
$B(n,k)(1+T_{\Ccal}(n,k,s_0))$ arithmetic operations.
It makes at most $B(n,k)$ calls, each with $k$ variables and at most
$s_0$ linear constraints.
\end{theorem}

\begin{proof}
On the region~$R$, every chosen affine term is no larger than any other
term for the same pair. By \Cref{lem:affine}, $D(x)$ is therefore exactly
the shortest-path metric of~$G_x$. Thus the subroutine receives a metric throughout $R$, and every
solution it finds gives a valid reweighting.
Conversely, any solution has an editable set $F$ of size~$k$ containing
all its changed edges. Its vector~$x$ realizes one of the enumerated sign
conditions. Every chosen term for that condition is minimal at~$x$, so
$x\in R$ and the corresponding subroutine call accepts.

There are at most $m^k\le n^{2k}$ choices of~$F$. The constants take
$2^kk!\poly(n)$ operations per choice, and $\Qcal$ contains at most
$n^24^k$ affine polynomials. \Cref{fact:bpr} enumerates its realizable
sign conditions in $2^{\bigoh(k^2)}n^{\bigoh(k)}$ operations.
Constructing the matrices and regions for all calls takes the same order of time.
The number of constraints in~\eqref{eq:region} is at most~$s_0$.
\end{proof}

We also mention that \cref{thm:template} can be analyzed in terms of bit-operations if all input weights are rationals.
To this end, let $Q$ be the product of the input edge-weight
denominators. Its bit length is $\bigoh(L)$, as is that of each integer
$Qw(e)$. Compute distances in $G-F$ with these integer weights and
divide the resulting finite constants by $Q$. Each finite $Qc_{uv,S}$
sums at most $(k+1)(n-1)$ scaled edge weights, so its bit length is
$\bigoh(L+\log n)$ after capping $k$ at $m$. Infinite terms are omitted from the comparisons. Multiplying each finite affine
comparison by $Q$ gives an integer polynomial whose coefficients have the same bit-length bound
without changing its sign. Thus the distance computations, comparisons,
and enumeration of regions use $B(n,k)\poly(L)$ bit operations by
\Cref{fact:bpr}. The affine distances passed to the target test are in
the original units.

\section{Reweighting for classes defined by local polynomial tests}\label{sec:local}

In this section, we use \cref{thm:template} to show \cref{thm:local}.
We then also show how to apply \cref{thm:local} to different target metrics.
Fix a local polynomial description $(r,\Phi_1,\ldots,\Phi_r)$ of
$\Ccal$. For each region supplied by \Cref{thm:template}, we test the
restrictions of the distance matrix to at most $r$ vertices using a
single vector of replacement weights.

\localTheorem*

\begin{proof}
Let $\Delta\ge1$ bound the degrees of the polynomials in the fixed tests.

Consider a call $(D(x),R)$ from \Cref{thm:template}. Fix a linear order
on the vertex set. For each nonempty $U\subseteq V$ with
$t=|U|\le r$, substitute the $\binom t2$ affine entries of $D(x)|_U$
into $\Phi_t$, using the induced order on~$U$. Denote the resulting
formula by $\Phi_U(x)$. The subroutine decides
\begin{equation}\label{eq:local-system}
  \exists x\in\Rbb^k\quad
  x\in R\ \land\!
  \bigwedge_{\substack{\varnothing\ne U\subseteq V\\ |U|\le r}}
  \Phi_U(x).
\end{equation}
A single vector of replacement weights must satisfy the tests for every subset.

For every $x\in R$, the matrix $D(x)$ is a metric. The formula for a
subset is therefore true exactly when its restricted metric belongs to
$\Ccal$. The locality assumption makes the conjunction equivalent to
$D(x)\in\Ccal$. Thus \eqref{eq:local-system} decides whether the region contains a feasible reweighting, and \Cref{thm:template} proves correctness.

There are at most $\sum_{t=1}^r\binom nt=O_r(n^r)$ subsets. Affine
substitution does not increase polynomial degree, so the formula has
$s+O_{\Ccal}(n^r)$ polynomial comparisons of degree at most $\Delta$, where
$s$ is the number of constraints defining~$R$. Its length has the same
bound. Each substituted polynomial can be expanded in
$(k+1)^{O_{\Ccal}(1)}$ arithmetic operations. By \Cref{fact:bpr},
feasibility can be decided within
\[
 \bigl((s+O_{\Ccal}(n^r))\Delta+2\bigr)^{\bigoh(k)}
 (k+1)^{O_{\Ccal}(1)}
\]
operations, including evaluation of the Boolean formula on the
realizable sign conditions. With $s\le k+\binom n2 2^k$, summing this cost over the calls in \Cref{thm:template} gives the asserted bound.

For rational graph weights, all affine coefficients have bit length
polynomial in $n$ and~$L$. The polynomials in the local tests have fixed degrees and fixed
rational coefficients. After substitution and clearing denominators,
their integer coefficients have bit length polynomial in $n$ and~$L$. Choosing positive common denominators preserves the signs
of the comparisons and does not require rescaling the metric. The bit version of \Cref{fact:bpr} gives the stated
bit bound and an exact real algebraic sample. Retaining the original
weights off~$F$ and using that sample on~$F$ returns a solution. Edges
whose sampled weights equal their original weights are not counted as
edited. The one-point case is decided directly by $\Phi_1$, and the
zero-edit case evaluates the same tests without real variables.
\end{proof}

\begin{corollary}\label{cor:linear}
Under the hypotheses of \Cref{thm:local}, if every polynomial compared with zero is
affine in the distances, every feasible rational input has a solution
with rational replacement weights. Such a solution can be returned within
the same bit bound.
\end{corollary}

\begin{proof}
After substitution all comparisons, including those defining $R$, are
affine. For each realizable sign condition that satisfies the Boolean
formula, its signs give a rational linear system. To obtain a rational
point when some inequalities are strict, introduce a common slack
variable $0\le\varepsilon\le1$, replace each strict inequality
$a^Tx<b$ by $a^Tx\le b-\varepsilon$, and maximize $\varepsilon$.
A point satisfying the original system has positive slack in every
strict inequality, so the optimum is positive. Conversely, any solution
with $\varepsilon>0$ satisfies all the original inequalities.
A polynomial-time rational linear programming algorithm returns an
optimal point of polynomial bit length when the optimum is positive.
Its running time is polynomial in the binary length of the entire
linear system, including the coefficient bits. If the system has no strict inequalities,
ordinary rational linear feasibility suffices. Enumerating the sign conditions and solving these linear programs takes
time within the stated bound.
\end{proof}

We next apply \cref{thm:local} to different target metrics.
For each class, it suffices to test subsets of at most $r$ points.
For a $t$-point metric with $1\le t\le r$, write $D_{ij}$ for the distance
between labels $i$ and $j$. Conditions on ordered triples or quadruples
range over all distinct labels and are vacuous if there are too few labels.

\paragraph*{Tree metrics and bounded Gromov hyperbolicity ($r=4$).}
For an ordered quadruple $(i,j,k,\ell)$, write
$A=D_{ij}+D_{k\ell}$, $B=D_{ik}+D_{j\ell}$, and $C=D_{i\ell}+D_{jk}$.
In the four-point definition, hyperbolicity at most $\delta$ means that
the two largest sums differ by at most $2\delta$. Thus for a fixed
rational $\delta\ge0$ we require
$(A\le B+2\delta)\lor(A\le C+2\delta)$ for every ordering.
At $\delta=0$ the maximum of the three sums must be attained at least
twice. By Buneman's four-point theorem~\cite[Theorem~2]{Buneman1974}, this is
equivalent to $D$ being a tree metric, that is, a metric realized by
distances in a tree with nonnegative edge lengths. The tree may contain
unlabeled vertices.

\paragraph*{The rectilinear plane ($r=6$).}
Write $\ell_1^d$ for $\Rbb^d$ with distance
$\|p-q\|_1=\sum_{h=1}^d |p_h-q_h|$. By Bandelt and Chepoi's six-point
theorem~\cite[Theorem~A]{BandeltChepoi1996}, a finite metric embeds
isometrically in $\ell_1^2$ if and only if every subset of at most six
points does. For $t\le6$, \Cref{lem:norm-tests} with $d=2$ supplies the
local test.

\begin{lemma}\label{lem:norm-tests}
For fixed positive integers $t,d$, the distance arrays of $t$ points
that admit a realization in $\ell_1^d$ are
described by a fixed Boolean combination of rational linear comparisons
in their $\binom t2$ entries. The formulas can be constructed effectively.
\end{lemma}
\begin{proof}
For $\ell_1^d$, introduce coordinates $p_{ih}$ and enumerate one ordering
of the $t$ points for each of the $d$ coordinates. There are $(t!)^d$
choices. Impose each ordering by weak inequalities, so ties are allowed.
For a fixed choice, the coordinate orders determine a linear expression
for each absolute difference, so
\[
 z_{ij}=\sum_{h=1}^d|p_{ih}-p_{jh}|
\]
is a linear equation in the coordinates and distance entries.
Fourier--Motzkin elimination removes all $td$ coordinates from the
resulting rational linear system. Projection of a polyhedron is a
polyhedron, so this produces a finite conjunction of rational linear
comparisons in the $z_{ij}$. Taking the disjunction over the coordinate
orders gives the required fixed formula.
\end{proof}

For fixed $d\ge3$, a bound on the size of the subsets needed to test embeddability in $\ell_1^d$ would yield a reweighting algorithm by \Cref{lem:norm-tests,thm:local}. See~\cite{BandeltChepoiLaurent1998} for the embedding question.

\paragraph*{Ultrametrics ($r=3$).}
For each ordered triple $(i,j,k)$, the strong triangle inequality is
$(D_{ij}\le D_{ik})\lor(D_{ij}\le D_{kj})$. This gives the required reweighting algorithm. 

However, note that Ultrametric reweighting is even fixed-parameter tractable in $k$.
An internal vertex $z$ on a shortest $u$-$v$-path would give
$D_{uv}=D_{uz}+D_{zv}>\max\{D_{uz},D_{zv}\}$.
Thus $G$ must be complete, and reweighting is exactly ultrametric editing
of its edge-weight matrix, solvable in $O(9^k\cdot k\cdot n^2)$ time
by the constrained ultrametric editing algorithm of Fomin, Golovach,
and More~\cite[Theorem~2]{fomin_et_al:LIPIcs.IPEC.2026.29}.
Together, these examples give the following corollary.

\begin{corollary}\label{cor:metric-applications}
For ultrametrics, tree metrics, metrics of Gromov hyperbolicity at most $\delta$ for any fixed~$\delta > 0$, and metrics embeddable isometrically in the rectilinear plane, shortest-path reweighting takes $2^{O(k^2)}n^{O(k)}$ arithmetic operations.
On rational input it takes $2^{O(k^2)}n^{O(k)}\poly(L)$ bit operations
and returns rational replacement weights on yes-instances.
\end{corollary}

\begin{proof}
Each local test above uses only rational linear comparisons in the distances.
The result follows from \Cref{thm:local,cor:linear}.
\end{proof}

The constants in \Cref{thm:local} include the locality bound, the degrees,
and the formula lengths. If the target class itself depends on an input
dimension, a separate argument is needed to obtain constants independent
of that dimension. \Cref{sec:euclid} provides such an argument for
Euclidean metrics.

\section{A Gram matrix test for Euclidean reweighting in variable dimension}\label{sec:euclid}

We now turn to the Euclidean metric space~$\Rbb^d$ for any dimension~$d$.
Applying \Cref{thm:template} directly with Menger's finite characterization~\cite[Theorem~1]{BowersBowers2017}---which says that a finite metric embeds isometrically in $\Rbb^d$ if and only if every subset of at most $d+3$ points does---would result in a running time of the form~$f(k,d) \cdot n^{\bigoh(kd)}$ for some exponential function~$f$.
We avoid the exponential dependency on the target dimension~$d$ (both in the function~$f$ and the polynomial in~$n$) by leveraging the classic Gram
criterion~\cite{Schoenberg1935,YoungHouseholder1938}.
In order to state it, we require the following definitions.
Let~$D$ be any symmetric~$n\times n$ matrix with all diagonal entries being zero and all entries being nonnegative.
Let $B$ be the $(n-1)\times (n-1)$ matrix with entries~$B_{uv}=\tfrac12\bigl(D_{un}^2+D_{vn}^2-D_{uv}^2\bigr)$, for each~$u,v \in [n-1]$.
Intuitively, $B$ records the inner products that a realization of~$D$ would have after placing the~$n$th point at the origin.
The Gram criterion states that~$D$ describes the distances between~$n$ points in Euclidean space~$\Rbb^d$ if and only if~$B$ is positive semidefinite and has rank at most~$d$.

On each region supplied by \Cref{thm:template}, the entries of~$D(x)$ are polynomials of degree at most one in~$x$.
Hence, the entries of~$B(x)$ have degree at most two in~$x$ as~$B$ is defined using the squares of entries in~$D$.
Hence, checking whether~$B(x)$ is positive semidefinite and computing its rank require a little additional care. 
We express these two requirements using the characteristic polynomial of~$-B(x)$, which is~$\det(tI+B(x))=t^{n-1}+c_1(x)t^{n-2}+\cdots+c_{n-1}(x)$.
We first show that~$B(x)$ is positive semidefinite and has rank at most~$d$ if and only if~$c_j(x) \geq 0$ for all~$j \in [n-1]$ and, when $d<n-1$, $c_{d+1}(x) = 0$. For convenience, we drop the argument~$x$ and prove the statement for any symmetric matrix~$D$ (with only nonnegative entries and zeros on the main diagonal).

\begin{lemma}\label{lem:test}
A matrix~$D$ embeds isometrically in~$\Rbb^d$ if and only if~$c_j\geq 0$ for each~${j\in[n-1]}$ and if~$d<n-1$, then~$c_{d+1}=0$.
\end{lemma}

\begin{proof}
By the Gram criterion, it suffices to show that the conditions on the coefficients hold if and only if~$B$ is positive semidefinite and has rank at most~$d$~\cite{Schoenberg1935,YoungHouseholder1938}.
Recall that a real symmetric matrix is positive semidefinite if and only if all of its eigenvalues are nonnegative.
So let~$\lambda_1,\ldots,\lambda_{n-1}$ be the (not necessarily distinct) eigenvalues of~$B$.
Since~$B$ is symmetric, these eigenvalues are real numbers.
Moreover, we have~$\det(tI + B) = (t+\lambda_1) (t+\lambda_2) \ldots (t+\lambda_{n-1})$ by the definition of eigenvalues.
This shows that~$c_j$ is the sum of all products of exactly~$j$ eigenvalues with distinct indices as these are precisely the terms that involve~$t^{n-1-j}$.

If~$B$ is positive semidefinite, then all its eigenvalues are
nonnegative and therefore~$c_j\geq 0$ for each~$j\in[n-1]$.
In the other direction, assume that all coefficients~$c_j$ are nonnegative.
Then, $\det(tI+B)>0$ for each~$t>0$, since the leading term~$t^{n-1}$ is positive and all other terms are nonnegative.
Now assume towards a contradiction that~$B$ is not positive semidefinite.
Then, there exists a negative eigenvalue~$\lambda_i$.
This yields~$\det(tI+B) = 0$ for~$t = -\lambda_i > 0$ as~$t+\lambda_i = 0$, a contradiction.
Hence, $B$ is positive semidefinite.

It remains to analyze the rank.
Assume that~$B$ is positive semidefinite.
The rank of~$B$ is then the number of its positive eigenvalues.
If~$d < n-1$, then all summands of~$c_{d+1}$ are nonnegative.
Moreover, at least one of them is positive if and only if at least~$d+1$ eigenvalues are positive.
Thus, $c_{d+1}=0$ if and only if~$B$ has rank at most~$d$.
If~$d \geq n-1$, then the rank of~$B$ is at most~$d$ since~$B$ is a~$(n-1) \times (n-1)$~matrix.
\end{proof}

We next show that since each entry of~$B(x)$ is a polynomial of degree at most~$2$ in~$x$, it holds that each coefficient~$c_j(x)$ is a polynomial of degree at most~$2j$ in~$x$.
Moreover, we show how to compute these polynomials.

\begin{lemma}\label{lem:charpoly}
Each coefficient polynomial~$c_j(x)$ has degree at most~$2j$ in~$x$, and all these polynomials can be computed in~$n^{\bigoh(k)}$ arithmetic operations.
For a region supplied by~\Cref{thm:template} on rational input weights, this takes~$n^{\bigoh(k)}\poly(L)$ bit operations. In this case, each coefficient in each polynomial~$c_j(x)$ can be encoded with~$\poly(n)\poly(L)$~bits.
\end{lemma}

\begin{proof}
We first analyze the degree of each coefficient polynomial~$c_j(x)$.
Recall that each entry of~$B(x)$ is a polynomial of degree at most~$2$ in~$x$.
Thus, when computing the determinant, each term
contributing to the coefficient~$c_j(x)$ of~$t^{n-1-j}$ takes~$t$ from~$n-1-j$ positions (which are then on the main diagonal as these are the only entries involving~$t$) and an entry of~$B(x)$ from each of the remaining $j$ positions.
Since each such entry has degree at most~$2$ in~$x$, the coefficient~$c_j(x)$ has degree at most~$2j$.

We next show how to compute these coefficient polynomials.
We do so using Newton's identities~\cite{Macdonald1998}.
To this end, we first need to define the trace~$s_i(x)$ of~$B(x)^i$.
The trace~$s_i(x)$ is the sum of entries on the main diagonal of~$B(x)^i$.
We obtain the matrix~$B(x)^i$ and its trace~$s_i(x)$ for each~$i \in [n-1]$ by successive multiplication and
computing the trace after each step.
For each fixed~$x$, let~$\lambda_1,\ldots,\lambda_{n-1}$ be the (not necessarily distinct) eigenvalues of~$B(x)$.
It is a well-known fact that the trace~$s_i(x)$ is equivalently the sum of the $i$th powers of the eigenvalues of~$B(x)$.
Newton's identities~\cite{Macdonald1998} now give~$j\cdot c_j(x)=\sum_{i=1}^{j}(-1)^{i-1}c_{j-i}(x)s_i(x)$ for each~$j\in[n-1]$, where~$c_0(x) = 1$.
Since only coefficients with smaller indices occur on the right, we can compute~$c_1(x),\ldots,c_{n-1}(x)$ in this order.

The next step is to analyze the number of arithmetic operations.
We first compute~$B(x)^i$ and~$s_i(x)$ for each~$i \in [n-1]$.
Computing one matrix requires~$\bigoh(n^3)$ multiplications and additions of polynomials.
Since we do this~$\bigoh(n)$ times, we add and multiply polynomials~$\bigoh(n^4)$ times.
We then compute the coefficient polynomials~$c_j(x)$ for increasing values of~$j \in [n-1]$.
Here, each new polynomial requires~$\bigoh(n)$ additions and multiplications of polynomials.
Thus, the number of arithmetic operations needed is~$\bigoh(n^4)$ times the number of operations needed for adding or multiplying two polynomials.
The entries of~$B(x)^i$ have degree at most~$2i$, so each product in the formula for~$c_j(x)$ has degree at most~$2(j-i)+2i=2j$.
Thus, every polynomial in the computation has degree at most~$2n-2$.
Each polynomial of degree at most~$2n-2$ in~$x$ is a linear combination of at most~$(2n-1)^k$ monomials, since each of the~$k$ exponents is an integer between~$0$ and~$2n-2$.
We can add two such polynomials coefficientwise and multiply them by considering all pairs of monomials.
Both operations therefore take at most~$\bigoh((2n-1)^{2k})$ arithmetic operations.
Since we perform~$\bigoh(n^4)$ polynomial operations, we can compute all coefficient polynomials~$c_j(x)$ in~$\bigoh(4^k n^{4+2k}) = n^{\bigoh(k)}$ arithmetic operations.

For rational input weights, let~$Q$ be the product of the input edge-weight denominators, as in the bit analysis in \cref{sec:template}.
Let~$A(x)=2Q^2B(x)$.
The entries of~$QD(x)$ are affine polynomials with integer coefficients of~$\bigoh(L+\log n)$ bits.
The entries of~$A(x)$ therefore have integer coefficients of~$\bigoh(L+\log n)$ bits as well.
For a polynomial~$p$, let~$h(p)$ be the sum of the absolute values of its coefficients, and let~$H=\max\{1,\max_{u,v} h(A_{uv})\}$.
Each entry of~$A(x)$ has at most~$\binom{k+2}{2} \in \bigoh(n^4)$ monomials, so~$\log(1+H)=\bigoh(L+\log n)$.
The trace~$r_i(x)$ of~$A(x)^i$ is a sum of at most~$n^i$ products of~$i$ entries of~$A(x)$.
Note that for two polynomials~$p$ and~$q$, it holds that~$h(p \cdot q)\leq h(p)h(q)$.
Hence, we have~$h(r_i(x))\leq(nH)^i$.
Let~$a_j(x)$ denote the coefficient of~$t^{n-1-j}$ in~$\det(tI+A(x))$.
By induction on~$j$, each product in Newton's formula
for~$ja_j(x)$ satisfies~$h(a_{j-i}(x)r_i(x))\le (nH)^{j-i}(nH)^i=(nH)^j$.
Thus, the sum of the absolute values of the coefficients of any partial sum is at most~$j(nH)^j$.
This gives~$h(a_j)\le(nH)^j$.
Hence, every intermediate integer coefficient in each computation can be encoded with~$\bigoh(n(L+\log n)) \in \poly(n)\poly(L)$ bits.

Finally, we analyze the number of bit operations in the case of rational input weights.
Note that each term contributing to~$c_j(x)$ uses exactly~$j$ entries of~$B(x)$, so~$a_j(x)=(2Q^2)^j c_j(x)$.
We can therefore use~$(2Q^2)^j$ as the common denominator for the coefficients of~$c_j(x)$.
It has~$\bigoh(nL)$ bits since~$Q$ has~$\bigoh(L)$~bits. Together with the arithmetic bound above, this gives~$n^{\bigoh(k)}\poly(L)$ bit operations.
This concludes the proof.
\end{proof}

We can now combine \Cref{thm:template,lem:test,lem:charpoly} to prove \Cref{thm:main}.

\euclideanTheorem*
\begin{proof}
Let~$(G,w,k,d)$ be an instance of \probSPtoEuc.
We assume without loss of generality that~$n \geq 1$ and~$1 \leq k \leq m$.
We apply \Cref{thm:template}, so it only remains to give an upper bound for the function~$T_{\Rbb^d}(n,k,s)$.
To this end, let~$(D,R)$ be the tuple provided by \Cref{thm:template}.
Recall that the task is to test whether there exists~$x \in R$ such that~$D(x)$ embeds isometrically in~$\Rbb^d$ in~$T_{\Rbb^d}(n,k,s)$ time, where~$s \leq k+\binom{n}{2}2^k$ is the number of constraints defining~$R$.
We then compute the matrix~$B(x)$, where~$B_{uv}(x)=\tfrac{1}{2}\bigl(D_{un}^2(x)+D_{vn}^2(x)-D_{uv}^2(x)\bigr)$ and the coefficient polynomials~$c_j(x)$ of the characteristic polynomial of~$-B(x)$ which is~$\det(tI+B(x))=t^{n-1}+c_1(x)t^{n-2}+\cdots+c_{n-1}(x)$ using \Cref{lem:charpoly}.
By \Cref{lem:test} it now suffices to test whether there exists~$x \in R$ such that~$c_j(x) \geq 0$ for all~$j \in [n-1]$ and if~$d < n-1$, then~$c_{d+1}(x) = 0$.
We do this using \Cref{fact:bpr}.
If the test accepts, a feasible vector~$x\in R$ gives weights for the~$k$ selected edges.

The algorithm is correct due to \Cref{thm:template}, so it only remains to analyze the number of operations.
For each guessed edge set and each region~$R$, \Cref{lem:charpoly} computes the coefficients using $n^{\bigoh(k)}$ arithmetic operations or~$n^{\bigoh(k)} \poly(L)$ bit operations in the case of rational input weights.
For a region with $s$ constraints, the feasibility test above uses at most $s+n$ polynomials of degree at most~$2n-2$ (with linear region constraints).
Thus, we have~$T_{\Rbb^d}(n,k,s_0) \in n^{\bigoh(k)}+(((k+\binom{n}{2}2^k)+n)n)^{\bigoh(k)} \subseteq 2^{\bigoh(k^2)}n^{\bigoh(k)}$.
By \Cref{thm:template}, the total number of arithmetic operations is at most~$B(n,k)(1+T_{\Rbb^d}(n,k,k+\binom{n}{2}2^k)) \in 2^{O(k^2)}n^{O(k)}$.
For rational input weights, the bit analysis in \cref{sec:template} gives integer comparisons for~$R$, and the proof of~\Cref{lem:charpoly} gives integer polynomials with the same signs as the~$c_j(x)$.
In both cases, the coefficients have bit length polynomial in~$n$ and~$L$.
The bit version of \Cref{fact:bpr}, together with the enumeration bound in~\Cref{thm:template}, therefore gives the claimed number of bit operations.
On a yes-instance, \Cref{fact:bpr} allows us to return an exact real point~$x\in R$ satisfying the conditions of \Cref{lem:test} for an accepting region~$R$.
This concludes the proof.
\end{proof}

\section{W[1]-hardness of Euclidean reweighting}
\label{sec:hardness}

We show that \probSPtoEuc{} is \Wone-hard for the parameter~$k$ for every~$d \geq 1$.
To this end, we first show hardness for~$d=1$ on triangle-free graphs.
We then use the triangle-freeness to lift the hardness to any~$d \geq 1$ irrespective of whether~$d$ is a constant or part of the input.
Our hardness also holds if a Hamiltonian path is provided as part of the input and if all input weights are positive integers encoded in unary.

\begin{proposition}
    \label{prop:d1hardness}
    \probSPtoEuc{} is \Wone-hard parameterized by~$k$ even when~$d=1$, the input graph is triangle-free, all input weights are positive integers encoded in unary, and a Hamiltonian path is given as part of the input.
\end{proposition}

\begin{proof}[Proof of \Cref{prop:d1hardness}]
    We present a reduction from \textsc{Multicolored Independent Set} parameterized by solution size.
    In this problem, one is given a graph~$G=(V,E)$, where~$V$ is partitioned into~$\ell$ cliques (with vertex sets~$V_1,V_2,\ldots,V_\ell$), each of size~$\nu=\frac{n}{\ell}$, where~$n=|V|$.
    The question is whether~$G$ contains an independent set of size~$\ell$, that is, exactly one vertex from each set~$V_i$ is contained in the independent set.
    This problem is known to be W[1]-hard~\cite{CyganFKLMPPS15}.
    We assume without loss of generality that~$\nu > 10\ell$.

    Before we present the reduction and formally prove its correctness, we first give a high-level idea.
    The construction will consist of two parts, a gadget for each set~$V_i$ and a construction to combine the different gadgets.
    The gadget for each~$V_i$ will be a cycle, where all edges have unit length.
    In order to embed the cycle on a line, at least one of the weights in the cycle has to be changed.
    We will enforce that there is precisely one edit in each cycle and this forces the changed weight to be a sufficiently large value.
    The modified edge will encode which vertex from the respective set~$V_i$ we pick in our solution independent set.
    We then also add relatively long edges between the different gadgets whenever there is an edge between the two respective endpoints in the input graph.
    We then ensure that the edges are longer than the distance between their endpoints in an embedding unless the edge has the very first vertex of the first gadget and the very last vertex of the second gadget as endpoints, that is, there exists an edge between the two chosen vertices in the input graph.

    We next present the formal construction.
    Let~$V_i = \{v^i_1,v^i_2,\ldots,v^i_{\nu}\}$ for all~$i \in [\ell]$.
    For each~$v^i_j$, we create three vertices~$p^i_j, q^i_j$ and~$r^i_j$.
    Let~$\mathcal{P}_i = \{p^i_j,q^i_j,r^i_j \mid j \in [\nu]\}$ and~${\mathcal{P} = \bigcup_{i \in [\ell]}\mathcal{P}_i}$.
    We then add the edges~$\{p^i_j,q^i_j\}$ and~$\{q^i_j,r^i_j\}$ with weight~$1$ for each~$i \in [\ell]$ and~$j \in [\nu]$.
    We also add the edges~$\{r^i_j,p^i_{j+1}\}$ for each~$i \in [\ell]$ and each~$j \in [\nu-1]$ as well as~$\{r^i_\nu,p^i_1\}$ for each~$i \in [\ell]$.
    All these edges also have weight~$1$.
    Next, we add~$2\ell$ additional vertices~$x_1,y_1,x_2,y_2,\ldots,x_{\ell},y_{\ell}$.
    Let~$\mathcal{X}$ be the set of all these vertices.
    For each~$i \in [\ell]$ and each~$j \in [\nu]$, we add the edges~$\{x_i,q^i_j\}$ and~$\{p^i_j,y_i\}$ with weight~$3\nu$.
    Moreover, we add a path~$Q_i$ with~$3\ell$ inner vertices (and thus~$3\ell+1$ edges) between~$y_i$ and~$x_{i+1}$ for each~$i \in [\ell-1]$.
    Each edge in these paths has weight~$3\nu+1$, and all inner vertices are newly created and distinct for different paths.
    We also add two vertices~$s$ and~$t$ and make~$s$ adjacent to (only)~$x_1$ and~$t$ to (only)~$y_{\ell}$ with edges of weight~$1$.
    Finally, for each edge~$\{v^{i_1}_{j_1},v^{i_2}_{j_2}\} \in E$ with~$i_1 < i_2$, we add an edge~$\{q^{i_1}_{j_1},p^{i_2}_{j_2}\}$ with weight~$(i_2-i_1+1)(3\nu-1)+(i_2-i_1)(3\nu+1)(3\ell+1)$.
    Setting~$k=3\ell$ concludes the construction.
    See \Cref{fig:construction} for an example and note that the constructed graph is triangle free by construction.
    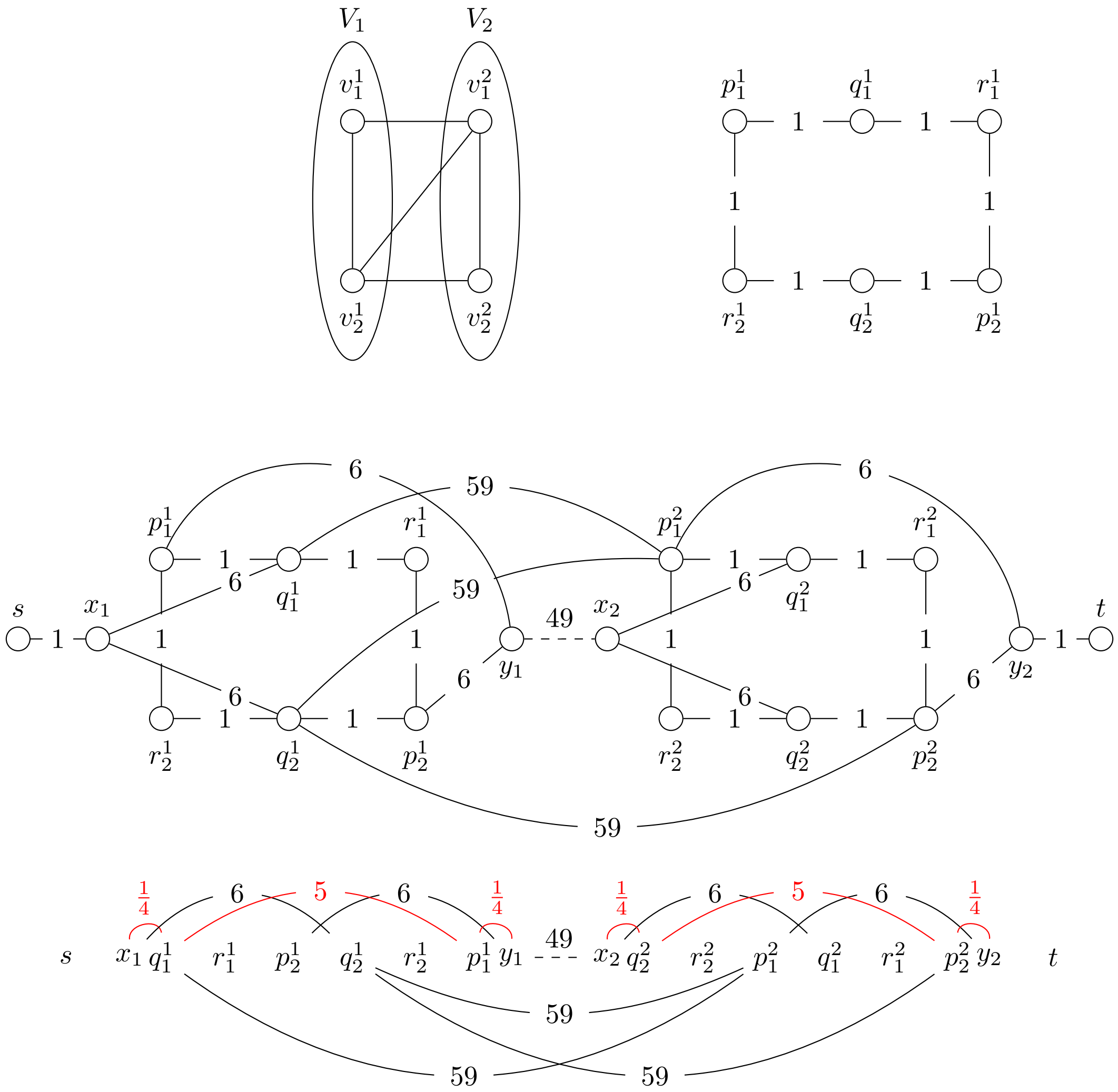
\begin{figure}[tp]
        \centering
        \begin{tikzpicture}[xscale=.8]
            \node[circle,draw,inner sep=3pt,label=$v^1_1$] at(-4,6.5) (v11) {};
            \node[circle,draw,inner sep=3pt,label=below:$v^1_2$] at(-4,4.5) (v12) {};
            \node[circle,draw,inner sep=3pt,label=$v^2_1$] at(-2,6.5) (v21) {};
            \node[circle,draw,inner sep=3pt,label=below:$v^2_2$] at(-2,4.5) (v22) {};
            \draw (v11) to (v21);
            \draw (v12) to (v21);
            \draw (v12) to (v22);
            \draw (v11) to (v12);
            \draw (v21) to (v22);
            \node[ellipse,draw,label=$V_1$,minimum height=4cm,minimum width=1cm] at(-4,5.5) {};
            \node[ellipse,draw,label=$V_2$,minimum height=4cm,minimum width=1cm] at(-2,5.5) {};

            \node[circle,draw,inner sep=3pt,label=$p^1_1$] at(2,6.5) (pp11) {};
            \node[circle,draw,inner sep=3pt,label=$q^1_1$] at(4,6.5) (qq11) {};
            \node[circle,draw,inner sep=3pt,label=$r^1_1$] at(6,6.5) (rr11) {};
            \node[circle,draw,inner sep=3pt,label=below:$p^1_2$] at(6,4.5) (pp12) {};
            \node[circle,draw,inner sep=3pt,label=below:$q^1_2$] at(4,4.5) (qq12) {};
            \node[circle,draw,inner sep=3pt,label=below:$r^1_2$] at(2,4.5) (rr12) {};
            \foreach \j in {1,2}{
                \draw (pp1\j) to node[circle,fill=white] {$1$}(qq1\j);
                \draw (qq1\j) to node[circle,fill=white] {$1$}(rr1\j);
            }
            \draw (pp11) to node[circle,fill=white] {$1$} (rr12);
            \draw (pp12) to node[circle,fill=white] {$1$} (rr11);

            \node[circle,draw,inner sep=3pt,label=$p^1_1$] at(-7,1) (p11) {};
            \node[circle,draw,inner sep=3pt,label=below:$q^1_1$] at(-5,1) (q11) {};
            \node[circle,draw,inner sep=3pt,label=$r^1_1$] at(-3,1) (r11) {};
            \node[circle,draw,inner sep=3pt,label=below:$p^1_2$] at(-3,-1) (p12) {};
            \node[circle,draw,inner sep=3pt,label=below:$q^1_2$] at(-5,-1) (q12) {};
            \node[circle,draw,inner sep=3pt,label=below:$r^1_2$] at(-7,-1) (r12) {};
            
            \node[circle,draw,inner sep=3pt,label=$s$] at(-9.25,0) (s) {};            \node[circle,draw,inner sep=3pt,label=$x_1$] at(-8,0) (x1) {};
            \node[circle,draw,inner sep=3pt,label=below:$y_1$] at(-1.5,0) (y1) {};  
            \node[circle,draw,inner sep=3pt,label=$x_2$] at(0,0) (x2) {};            \node[circle,draw,inner sep=3pt,label=below:$y_2$] at(6.5,0) (y2) {};
            \node[circle,draw,inner sep=3pt,label=$t$] at(7.75,0) (t) {};
            
            \node[circle,draw,inner sep=3pt,label=$p^2_1$] at(1,1) (p21) {};
            \node[circle,draw,inner sep=3pt,label=below:$q^2_1$] at(3,1) (q21) {};
            \node[circle,draw,inner sep=3pt,label=$r^2_1$] at(5,1) (r21) {};
            \node[circle,draw,inner sep=3pt,label=below:$p^2_2$] at(5,-1) (p22) {};
            \node[circle,draw,inner sep=3pt,label=below:$q^2_2$] at(3,-1) (q22) {};
            \node[circle,draw,inner sep=3pt,label=below:$r^2_2$] at(1,-1) (r22) {};

            \foreach \i in {1,2}{
                \foreach \j in {1,2}{
                    \draw (p\i\j) to node[circle,fill=white] {$1$}(q\i\j);
                    \draw (q\i\j) to node[circle,fill=white] {$1$}(r\i\j);
                }
            \draw (p\i1) to node[circle,fill=white] {$1$} (r\i2);
            \draw (p\i2) to node[circle,fill=white] {$1$} (r\i1);
            }

            \draw (s) to node[midway,circle,fill=white,inner sep=1pt] {$1$} (x1);
            \draw (x1) to node[near end,circle,fill=white,inner sep=1pt] {$6$} (q11);
            \draw (x1) to node[near end,circle,fill=white,inner sep=1pt] {$6$} (q12);
            \draw[bend right=70] (y1) to node[circle,fill=white] {$6$} (p11);
            \draw (y1) to node[circle,fill=white] {$6$} (p12);
            \draw[dashed] (y1) to node[above, midway,circle,fill=white,inner sep=1pt] {$49$} (x2);
            \draw (x2) to node[near end,circle,fill=white,inner sep=1pt] {$6$} (q21);
            \draw (x2) to node[near end,circle,fill=white,inner sep=1pt] {$6$} (q22);
            \draw[bend right=70] (y2) to node[circle,fill=white] {$6$} (p21);
            \draw (y2) to node[circle,fill=white] {$6$} (p22);
            \draw[bend left=30] (q11) to node[circle,fill=white] {$59$} (p21);
            \draw[bend left=20] (q12) to node[circle,fill=white] {$59$} (p21);
            \draw[bend right=27] (q12) to node[circle,fill=white] {$59$} (p22);
            \draw (y2) to node[midway,circle,fill=white,inner sep=1pt] {$1$} (t);
            
            \node (S) at(-8.5,-4) {$s$};
            \node (X1) at(-7.5,-4) {$x_1$};
            \node (Q11) at(-7,-4) {$q^1_1$};
            \node (R11) at(-6,-4) {$r^1_1$};
            \node (P12) at(-5,-4) {$p^1_2$};
            \node (Q12) at(-4,-4) {$q^1_2$};
            \node (R12) at(-3,-4) {$r^1_2$};
            \node (P11) at(-2,-4) {$p^1_1$};
            \node (Y1) at(-1.5,-4) {$y_1$};
            \node (X2) at(0,-4) {$x_2$};
            \node (Q22) at(.5,-4) {$q^2_2$};
            \node (R22) at(1.5,-4) {$r^2_2$};
            \node (P21) at(2.5,-4) {$p^2_1$};
            \node (Q21) at(3.5,-4) {$q^2_1$};
            \node (R21) at(4.5,-4) {$r^2_1$};
            \node (P22) at(5.5,-4) {$p^2_2$};
            \node (Y2) at(6,-4) {$y_2$};
            \node (T) at(7,-4) {$t$};
            
            \draw[bend left=90,red] (X1) to node[above] {$\frac{1}{4}$} (Q11);
            \draw[bend left=40] (X1) to node[circle,fill=white] {$6$} (Q12);
            \draw[bend left=40] (P12) to node[circle,fill=white] {$6$} (Y1);
            \draw[bend left=40] (X2) to node[circle,fill=white] {$6$} (Q21);
            \draw[bend left=30,red] (Q11) to node[circle,fill=white] {$5$} (P11);
            \draw[bend left=90,red] (P11) to node[above] {$\frac{1}{4}$} (Y1);
            \draw[bend left=90,red] (X2) to node[above] {$\frac{1}{4}$} (Q22);
            \draw[bend left=40] (P21) to node[circle,fill=white] {$6$} (Y2);
            \draw[bend left=30,red] (Q22) to node[circle,fill=white] {$5$} (P22);
            \draw[bend right=30] (Q11) to node[circle,fill=white] {$59$} (P21);
            \draw[bend right=20] (Q12) to node[circle,fill=white] {$59$} (P21);
            \draw[bend right=30] (Q12) to node[circle,fill=white] {$59$} (P22);
            \draw[bend left=90,red] (P22) to node[above] {$\frac{1}{4}$} (Y2);
            \draw[dashed] (Y1) to node[above] {$49$} (X2);
        \end{tikzpicture}
        \caption{The top left shows an input instance of \textsc{Multicolored Independent Set} with~$\ell=2$ and~${\nu=2}$. The top right shows the gadget constructed for~$V_1$ and the middle shows the entire constructed instance of \textsc{\probSPtoEuc}, where the path~$Q_1$ between~$y_1$ and~$x_2$ is only indicated by a dashed edge. The bottom shows a solution corresponding to the independent set~$\{v_1^1,v^2_2\}$ with~$k=6$ weight modifications. The distance between two consecutive displayed points on the line is always~$1$ and the respective edge is not drawn.
        The only exceptions are the edges~$\{x_1,q_1^1\}, \{p_1^1,y_1\}, \{x_2,q^2_2\}, \{p^2_2,y_2\}$ which have distance~$\frac{1}{4}$, and the gap between~$y_1$ and~$x_2$, which has length~$49$ and contains the omitted inner vertices of~$Q_1$. The solution changes weights of the edges~$\{x_1,q_1^1\},\{p^1_1,q^1_1\},\{p_1^1,y_1\},\{x_2,q_2^2\},\{p_2^2,q^2_2\}$, and~$\{p^2_2,y_2\}$ drawn in red.}
        \label{fig:construction}
    \end{figure}

    Note that the constructed graph contains a Hamiltonian path as follows.
    Starting from~$s$, visit, for increasing values of~$i$ from~$1$ to~$\ell$, the vertex~$x_i$, all vertices of~$\mathcal{P}_i$ along the cycle with~$\{p_1^i,q_1^i\}$ omitted (from~$q_1^i$ to~$p_1^i$), and the vertex~$y_i$.
    Between consecutive gadgets, it follows~$Q_i$ from~$y_i$ to~$x_{i+1}$, and after~$y_\ell$ it ends at~$t$.
    We now prove that the construction is correct.
    Since the construction can be computed in polynomial time and all constructed numbers are positive integers that are polynomial in the input size, this will conclude the proof.
    Assume first that there is an independent set~$\mathcal{I}$ of size~$\ell$ in~$G$.
    We show how to modify~$k=3\ell$ edge weights such that the resulting graph can be embedded on a line.
    For each~$i\in [\ell]$, let~$v^i_{j_i}$ be the vertex in~$V_i \cap \mathcal{I}$.
    We change the weight of the edge~$\{p^i_{j_i},q^i_{j_i}\}$ to~$3\nu-1$ for each~$i \in [\ell]$.
    Moreover, for each~$i\in [\ell]$, we change the weight of the edges~$\{x_{i},q^{i}_{j_i}\}$ and~$\{p^i_{j_i},y_i\}$ to~$\frac{1}{2\ell}$.
    Note that we changed exactly~$3\ell$ edge weights and no edge weight is set to~$0$.
    We next show how to embed the vertices isometrically onto a line.
    
    We place all vertices from left to right in the following sequence, where the distance between two consecutive vertices in the sequence is~$1$ unless it is an edge in one of the paths~$Q_i$ for some~$i \in [\ell-1]$ or an edge between~$x_i$ or~$y_i$ and a vertex in~$\mathcal{P}_i$.
    In the former case, the distance will be~$3\nu+1$ (the weight of the original edge) and in the latter case, the distance will be~$\frac{1}{2\ell}$.
    The sequence is~$(s,x_1) \circ S_1 \circ Q_1 \circ S_2 \circ \ldots \circ Q_{\ell-1} \circ S_{\ell} \circ (y_{\ell},t)$, where~$\circ$ is the concatenation of sequences, ${S_i = (q^i_{j_i},r^i_{j_i},p^i_{j_i+1},q^i_{j_i+1},\ldots,r^i_{\nu},p^i_1,q^i_1,r^i_1,p^i_2,\ldots,r^i_{j_i-1},p^i_{j_i})}$ for each~$i \in [\ell]$, and~$Q_i$ is directed from~$y_i$ to~$x_{i+1}$. 
    Note that for each consecutive pair of vertices, there is an edge with the described length between the two vertices.

    It remains to verify that no edge weight is shorter than the distance in the constructed embedding.
    Every unmodified cycle edge joins consecutive vertices of its sequence~$S_i$ and its endpoints therefore have distance~$1$.
    Every edge in any path~$Q_i$ for any~$i \in [\ell-1]$ as well as the two edges incident to~$s$ and~$t$ also have weight equal to the distance between their endpoints.
    The weight of edges~$\{p^i_{j_i},q^i_{j_i}\}$ for~$i \in [\ell]$ was changed to~$3\nu-1$, which is the distance in the embedding of~$S_i$.
    All edges~$\{x_{i},q^i_j\}$ and~$\{p^i_j,y_i\}$ for~$i\in [\ell]$ and~$j \in [\nu]$ are either of weight~$3\nu$ and therefore longer than the distance between their respective endpoints in the embedding or are changed to~$\frac{1}{2\ell}$, in which case the two respective endpoints also have distance~$\frac{1}{2\ell}$ in the embedding.
    Finally, we consider edges of the form~$\{q^{i_1}_{j_1},p^{i_2}_{j_2}\}$.
    The maximum distance in the embedding between a vertex in~$\mathcal{P}_{i_1}$ and in~$\mathcal{P}_{i_2}$ is~$(i_2-i_1+1)(3\nu-1) + (i_2-i_1) ((3\ell+1)(3\nu+1)+\frac{1}{\ell})$.
    Since~$i_2-i_1 < \ell$, the distance is larger than~$(i_2-i_1+1)(3\nu-1) + (i_2-i_1)(3\ell+1)(3\nu+1)$ (the weight of the respective edge) only if~$q^{i_1}_{j_1}$ is the very first vertex in~$S_{i_1}$ and~$p^{i_2}_{j_2}$ is the very last vertex in~$S_{i_2}$.
    This cannot be as this would imply that~$\{v^{i_1}_{j_1},v^{i_2}_{j_2}\} \in E$, a contradiction to the fact that~$\mathcal{I}$ is an independent set and both~$v^{i_1}_{j_1}$ and~$v^{i_2}_{j_2}$ are contained in~$\mathcal{I}$.
    Hence, the constructed embedding is isometric.

    In the other direction, assume that an isometric embedding onto the line exists after changing at most~$k=3\ell$ edge weights.
    Note that all edge weights remain positive and hence distinct vertices are embedded at distinct positions.
    Note that any two consecutive vertices in the embedding are adjacent and their edge has weight equal to their distance.
    Let~$P$ be the Hamiltonian path obtained by ordering all vertices according to their positions on the line.
    Since~$s$ and~$t$ have degree one, they are the endpoints of~$P$.
    Every inner vertex of each of the paths~$Q_i$ for each~$i \in [\ell-1]$ has degree two and since it cannot be an endpoint of~$P$, both incident edges belong to~$P$.
    Thus, every edge of every path~$Q_i$ belongs to~$P$, as do the edges~$\{s,x_1\}$ and~$\{y_\ell,t\}$.

    We next show that the weight of at least one cycle edge in~$\mathcal{P}_i$ has to be modified for each~$i\in[\ell]$.
    Assume towards a contradiction that all edges of one such cycle retain weight~$1$.
    Since~$3\nu>2k$ and at most~$2k$ vertices are incident to modified edges, there is a vertex~$u$ on this cycle that is not incident to any modified edge.
    All edges incident to~$u$ have weight at least~$1$, so every other vertex has distance at least~$1$ from~$u$.
    Its two neighbors on the cycle have distance exactly~$1$ from~$u$ and therefore lie on opposite sides of~$u$ in the embedding.
    The path between these two neighbors obtained by deleting~$u$ from the cycle must consequently contain an edge whose endpoints lie on opposite sides of~$u$ and are therefore at distance at least~$2$.
    This contradicts the fact that all edges in this path have weight~$1$.

    Next, we show that for each~$i\in[\ell]$, at least one edge between~$x_i$ and~$\mathcal{P}_i$ and at least one edge between~$y_i$ and~$\mathcal{P}_i$ have to be modified.
    Assume towards a contradiction that none of the edges~$\{x_i,q_j^i\}$ is modified.
    A shortest path from~$x_i$ to any of these neighbors either starts with one of these edges and thus already has length at least~$3\nu$, or leaves~$x_i$ along the path~$Q_{i-1}$ if~$i>1$---for~$i=1$, the only other edge is towards~$s$ and is not part of any shortest path to a vertex in~$\mathcal{P}_i$.
    If the shortest path enters~$Q_{i-1}$, then it must traverse the entire path before reaching a vertex in~$\mathcal{P}_i$.
    This connecting path has~$3\ell+1=k+1$ edges, so at least one of them retains weight~$3\nu+1$.
    Such a path therefore has length greater than~$3\nu$.
    It follows that every~$q_j^i$ has distance exactly~$3\nu$ from~$x_i$.
    This is impossible, since there are~$\nu>2$ such vertices but at most two positions on the line at distance~$3\nu$ from~$x_i$.
    The same argument applies to~$y_i$ and its neighbors~$p_j^i$.
    The~$\ell$ required cycle modifications and the~$2\ell$ required modifications of edges between~$\mathcal{X}$ and~$\mathcal{P}$ are all distinct and exhaust the budget~$k=3\ell$.
    Hence, exactly one cycle edge in each~$\mathcal{P}_i$ is modified, exactly one edge between each vertex in~$\mathcal{X}$ and its respective cycle is modified, and no other edge weight changes.

    We next show that~$P$ does not use any edge between two different sets~$\mathcal{P}_{i_1}$ and~$\mathcal{P}_{i_2}$.
    Assume towards a contradiction that~$P$ contains such an edge~$e=\{u,v\}$.
    As argued above, this edge is unmodified and therefore has weight at least~$6\nu$.
    Since~$u$ and~$v$ are consecutive in the embedding, their distance is also at least~$6\nu$.
    Note that deleting all modified edges and all edges between different cycle gadgets leaves the graph connected.
    Hence, there is a path from~$u$ to~$v$ using only unmodified edges of weight at most~$3\nu+1$.
    Since no vertex is embedded strictly between~$u$ and~$v$, this path must contain an edge whose endpoints lie on opposite sides of the gap between them.
    The distance between these endpoints is at least~$6\nu$, contradicting the fact that their edge has weight at most~$3\nu+1<6\nu$.

    We may assume without loss of generality that~$s$ is embedded to the left of~$t$.
    After deleting all edges between different cycle gadgets (which are not part of~$P$ as shown above), every edge of~$Q_i$ separates the first~$i$ gadgets from the remaining gadgets.
    It follows that~$P$ visits the gadgets in increasing order of their indices.
    Moreover, between~$x_i$ and~$y_i$, the path~$P$ must visit all vertices of~$\mathcal{P}_i$ consecutively, since these two vertices are the only attachments of that cycle to the rest of the remaining graph.
    The part of~$P$ within~$\mathcal{P}_i$ is therefore a spanning path of its cycle and omits exactly one cycle edge.
    Let~$e_i$ be this omitted edge.
    Note that the weight of~$e_i$ must be modified to at least~$3\nu-1$.
    The leftmost vertex from each set~$\mathcal{P}_i$ is adjacent to~$x_i$ and hence is a vertex~$q^i_j$ for some~$j \in [\nu]$.
    The rightmost vertex is adjacent to~$y_i$ and hence is~$p^i_j$ for some~$j \in [\nu]$.
    Thus, $e_i=\{p_{j_i}^i,q_{j_i}^i\}$ for some~$j_i \in [\nu]$ for each~$i \in [\ell]$.
    
    To conclude the proof, we will show that~$\mathcal{I}=\{v_{j_i}^i\mid i\in[\ell]\}$ is a multicolored independent set.
    Assume towards a contradiction that there are~$i_1<i_2$ such that~$\{v_{j_{i_1}}^{i_1},v_{j_{i_2}}^{i_2}\}\in E$.
    By construction, the edge~$\{q_{j_{i_1}}^{i_1},p_{j_{i_2}}^{i_2}\}$ exists and has unchanged weight~$(i_2-i_1+1)(3\nu-1)+(i_2-i_1)(3\ell+1)(3\nu+1)$.
    Let~$\varepsilon>0$ be the minimum distance between two consecutive vertices in the embedding.
    Note that all vertices of each set~$\mathcal{P}_i$ for each~$i_1 \leq i \leq i_2$ and all vertices of each path~$Q_i$ with~$i_1 \leq i < i_2$ are embedded between~$q_{j_{i_1}}^{i_1}$ and~$p_{j_{i_2}}^{i_2}$.
    For each set~$\mathcal{P}_i$, the vertices are embedded consecutively and the maximum distance between any two vertices in it is exactly~$3\nu-1$.
    The paths~$Q_i$ are also embedded consecutively and span a distance of~$(3\ell+1)(3\nu+1)$ each.
    Moreover, the edge~$\{p^{i_1}_{j_{i_1}},y_{i_1}\}$ is also part of a shortest path between~$q_{j_{i_1}}^{i_1}$ and~$p_{j_{i_2}}^{i_2}$.
    The distance between these two vertices is therefore at least~$(i_2-i_1+1)(3\nu-1)+(i_2-i_1)(3\ell+1)(3\nu+1) + \varepsilon$, which is strictly greater than the weight of the edge.
    This contradicts the assumption that the embedding is isometric.
    Thus, $\mathcal{I}$ is an independent set containing exactly one vertex from each color class.
    This concludes the proof.
\end{proof}

We next prove \Cref{thm:euclidean-hardness}, which we restate here for convenience.

\euclideanHardnessTheorem*

\begin{proof}
    We present a reduction from \probSPtoEuc{} in a single dimension, which is \Wone-hard by \Cref{prop:d1hardness}.
    Let~$(G,w,k,1)$ be an input instance of \probSPtoEuc{} where~$G$ is connected and triangle-free, all weights of~$w$ are encoded in unary, and a Hamiltonian path~$P$ in~$G$ is known.
    We return the instance~$(G,w,k,d)$.
    It only remains to prove correctness.
    If~$(G,w,k,1)$ is a yes-instance, then~$(G,w,k,d)$ is also a yes-instance as~$d \geq 1$ and any embedding on a line is also an embedding in dimension~$d$.
    So assume for the other direction that~$(G,w,k,d)$ is a yes-instance of \probSPtoEuc.
    We will show that any isometric embedding of~$G$ only uses a single line as~$G$ is triangle-free.
    This then proves that~$(G,w,k,1)$ is also a yes-instance of \probSPtoEuc{} and concludes the proof.
    So assume towards a contradiction that the vertices of~$G$ are not embedded on a single line.
    Then, there are three vertices~$u,v$, and~$w$ that do not lie on the same line.
    We pick three such points that minimize the area of the triangle between them.
    By minimality, no other vertex is embedded in the triangle or on its sides, since together with two of its vertices it would form a nondegenerate triangle of strictly smaller area.
    No other vertex can be embedded on a corner of the triangle as no two vertices are embedded onto the same point.
    We claim that~$u,v$, and~$w$ must form a triangle.
    To this end, consider~$u$ and~$v$.
    The distance between their points must be equal to the length of a shortest path between them.
    Since no point is embedded between them, there must be an edge between~$u$ and~$v$.
    The argument for edges~$\{u,w\}$ and~$\{v,w\}$ is completely analogous.
    Thus, all three of these edges are present in~$G$, contradicting that~$G$ is triangle-free.
    This concludes the proof.
\end{proof}

\section{Conclusion}\label{sec:conclusion}

We have initiated the study of how changing a few edge weights can make
a graph's shortest-path metric admit an isometric embedding in a prescribed
space. A single edit can change the distances between many pairs of
vertices, but which pairs it affects depends on the routes available in
the graph. This raises the question of identifying restrictions on the graph
that can make it easier to find the edits. 

\paragraph*{Acknowledgments.}

The authors would like to thank Saket Saurabh for fruitful discussions on the topic. MB was supported by the German Research Foundation (DFG), SPP 2378: \mbox{ReNO-2} (project number 511099228). FF was supported by the Research Council of Norway under the BWCA project, reference 314528, and European Research Council (ERC) via grant NewPC, reference 101199930. PG was supported by the Research Council of Norway under the BWCA  (grant no.~314528) and Extreme-Algorithms (grant no~355137) projects.

\paragraph*{AI Statement.}
The authors developed an algorithm for the Euclidean problem and the $\Wone$-hardness result for the same metric (\Cref{thm:euclidean-hardness}) without any AI assistance.
The algorithm runs in $f(d,k)\cdot n^{\bigoh(kd)}$ time.
Conversations with \mbox{GPT-6} models  then led to an algorithm running in $2^{\bigoh(k^2)}n^{\bigoh(k)}$ arithmetic operations with constants independent of~$d$ (\Cref{thm:main}). Identifying which properties of the target space this algorithm actually uses led the authors to formulating \Cref{thm:local}.

The same model spotted a small change (the addition of long paths between the vertices~$y_i$ and~$x_{i+1}$) that enabled a simpler argument for one step in the proof for \Cref{thm:euclidean-hardness}: any solution must place all gadgets in the intended order.
The rest of the reduction and proof is as the authors originally developed it.

The authors also used generative AI to proofread the manuscript, to help with exposition and figures, to search the literature, to identify applications, and to check technical arguments.
The authors take full responsibility for the content and the writing of this work.

\bibliographystyle{plainnat}
\bibliography{bib.bib}

\end{document}